\documentclass{aa}  
\usepackage{graphicx}
\usepackage{txfonts}
\usepackage{graphicx}
\usepackage{lscape}
\usepackage{subfigure}
\usepackage{rotating}
\usepackage{array,longtable}
\usepackage{lscape}
\usepackage{multicol}
\newcommand{\micron}{{\rm \mu m}}

\begin{document} 

\title{A CO observation of the galactic methanol masers}


\author{Zhiyuan Ren\inst{1,2}\thanks{E-mail:renzy@bao.ac.cn} \and Yuefang Wu\inst{1}\thanks{E-mail:yfwu@pku.edu.cn}  \and Tie Liu\inst{1} \and Lixin Li\inst{3} \and Di Li\inst{2} \and Binggang Ju$^{4,5}$}

\institute{
The Department of Astronomy, Peking University, No.5 Yiheyuan Road, HaiDian District, Beijing 100871, P. R. China
\and
National Astronomical Observatories (NAOC), Chinese Academy of Science, Chaoyang District, Datun Rd A20, Beijing, 100012, China;
\and
The Kavli Institute for Astronomy and Astrophysics (KIAA), Peking University (PKU), No.5 Yiheyuan Road, HaiDian District, Beijing 100871, P. R. China;
\and
Purple Mount Observatory, Chinese Academy of Science, PO Box 26, Delingha, Qinghai 817000, China
\and
State Key Laboratory of Radio Astronomy, Chinese Academy of Science
}

\date{Received mm-dd, 2012; accepted mm-dd, 2013}

 
  \abstract
   {We investigated the molecular gas associated with 6.7 GHz methanol masers throughout the Galaxy using a $J=1-0$ transition of the CO isotopologues.}
   {The methanol maser at 6.7 GHz is an ideal tracer for young high-mass star-forming cores. Based on molecular line emissions in the maser sources throughout the Galaxy, we can estimate their physical parameters and, thereby, investigate the forming conditions of the high-mass stars. }
   {Using the 13.7-meter telescope at the Purple Mountain Observatory (PMO), we have obtained ${\rm ^{12}CO}$ and ${\rm ^{13}CO}$ $(1-0)$ lines for 160 methanol masers sources from the first to the third Galactic quadrants. We made efforts to resolve the distance ambiguity by careful comparison with the radio continuum and HI 21 cm observations. We examined the statistical properties in three aspects: first, the variation throughout the Galaxy; second, the correlation between the different parameters; third, the difference between the maser sources and the infrared dark clouds. In addition, we have also carried out ${\rm ^{13}CO}$ mapping for 33 sources in our sample.}
   {First, the maser sources show increased $^{13}$CO line widths toward the Galactic center, suggesting that the molecular gas are more turbulent toward the Galactic center. This trend can be noticeably traced by the ${\rm ^{13}CO}$ line width. In comparison, the Galactic variation for the H$_2$ column density and the $^{12}$CO excitation temperature are less significant. Second, the $^{12}$CO excitation temperature shows a noticeable correlation with the H$_2$ column density. A possible explanation consistent with the collapse model is that the higher surface-density gas is more efficient to the stellar heating and/or has a higher formation rate of high-mass stars. Third, comparing the infrared dark clouds, the maser sources on average have significantly lower H$_2$ column densities, moderately higher temperatures, and similar line widths. Fourth, In the mapped regions around 33 masers, 51 $^{13}$CO cores have been revealed. Among them, only 17 coincide with the radio continuum emission ($F_{\rm cm}>6$ mJy), while a larger fraction (30 cores) coincide with the infrared emissions. Only one maser source has no significant IR emission. The IR-bright and radio-bright sources exhibit significantly higher $^{12}$CO excitation temperatures than the IR-faint and radio-faint sources, respectively. }
   {The 6.7 GHz masers show a moderately low ionization rate but have a common-existing stellar heating that generates the IR emissions. The relevant properties can be characterized by the ${\rm ^{12}CO}$ and ${\rm ^{13}CO}~(1-0)$ emissions in several aspects as described above. }


   \maketitle
%

\section{Introduction}
The high-mass (O and early B-type) stars play a critical role in the mass distribution and chemical evolution of the galaxies. Yet, their forming processes are still not fully understood. The main difficulty in revealing their birth is that, most young high-mass stars are deeply obscured by a dust envelope and therefore, hard to observe at optical wavelengths. Moreover, they are usually at large distances of several kilo parsecs and have a clustered environment.

Masers, such as CH$_3$OH masers~\citep{menten91,walsh98,beuther02}, were found to be an ideal tracer for the forming sites of young high-mass stars. In the targeted survey at early decades, the methanol masers were observed exclusively near the Ultra Compact (UC) H{\sc ii} regions \citep{menten92,sutton01} and totally absent in other types of objects, including the forming regions of low-mass stars, AGB star, and other galaxies. The subsequent observations \citep{phillips98,walsh98} have revealed a greater number of  6.7 GHz masers without a radio continuum counterpart and suggest that they could trace even earlier stages (prior to the UC H{\sc ii} stage). Based on the accumulated data and newly performed surveys, \citet{pestalozzi05} have presented a relatively completed catalogue of the 6.7 GHz methanol masers throughout the Galaxy. After that, more extensive molecular molecular line surveys \citep[][etc.]{minier05,purcell06,longmore06} were followed as a major procedure to investigate the physical and chemical properties in the methanol-maser sources. As a result, they are detected in many dense molecular tracers such as H$^{13}$CO$^+$, CH$_3$CN, and sulfur-bearing molecules. 

On the other hand, the different methanol maser transitions \citep[classified to be class-I and class-II][]{menten91} are found to trace different physical conditions. Besides the widely detected 6.7 GHz masers, which belong to class-II \citep{menten91}, various class-I masers were observed at the interface between the outflow and the ambient gas \citep[][etc.]{plambeck90,johnston97,voronkov05}. The latter observations with higher resolution showed that the Class-I masers arise from intense outflow and shock emissions \citep{cyganowski08,wu10}. These properties are consistent with the theoretical anticipation that the Class-I masers are stimulated by the collisional processes while the class-II transitions are mainly excited by the radiative processes \citep[][]{cragg92}.

Using the 13.7-m millimeter telescope at the Qinghai station of the Purple Mountain Observatory, \citet{liu10a} observed 98 maser sources (including both type I and II) and studied the relation between their CO-line emissions and infrared luminosities. Since the 13.7-m telescope is appropriate for single-point and mapping observations over a large area, we carried out a more extensive observation towards the Galactic methanol-maser sources, aiming to get a more complete sample of the molecular lines and thereby analyze their physical parameters and statistical properties.   

The targeted 6.7 GHz maser sources (referred to as "maser sources" hereafter) for the current observation are taken from the \citet{pestalozzi05}'s catalogue and mainly in the northern sky (within the Dec. range available for the telescope) in $^{12}$CO and $^{13}$CO $(1-0)$. We estimated the physical parameters from these lines, examined their Galactic distribution and correlations. In addition, we made ${\rm ^{13}CO}$ maps for 33 sources in order to study the gas distribution around the maser sources. We report the observation and data reductions in Section 2. We show the measurement of the physical parameters in Section 3. In Section 4, we discuss the statistical properties of the physical parameters and made a comparison with the physical parameters in the infrared dark clouds (IRDCs) that are sampled by \citet{du08}. A summary of the major results is presented in Section 5.

\section{Observation and data reduction}
The observations of the $J=1-0$ lines of the CO isotopologues were performed using the PMO 13.7-m millimeter telescope \citep{zuo04,zuo11} from September to December in 2009. A superconductor-insulator-superconductor (SIS) receiver with three acoustic-optical spectrometers (AOSs) working at 110.20 and 115.27 GHz covered ${\rm ^{13}CO}$ and ${\rm ^{12}CO}$ $(1-0)$ lines respectively. Every spectrometer had 1024 channels. The bandwidths for $^{12}$CO and $^{13}$CO lines were 145.330 and 42.762 MHz, corresponding to velocity resolution of $\Delta V=0.37$ and 0.11 km s$^{-1}$, respectively. The weather conditions had been good. The system temperature during observations was between 150 and 300 K. The rms noise level per channel was 0.4 and 0.3 K for the $^{12}$CO and $^{13}$CO bands, respectively. The full-width half-maximum (FWHM) beam width was $\sim57''$\citep{zuo11}. Two objects, W51D or S140 was observed once per hour as flux calibrators (one of them closer to the target source would be adopted). The main-beam brightness temperature of the calibrators ($T_{\rm mb,calibrator}$) were estimated from the recently released beam efficiency\footnote{http://www.dlh.pmo.cas.cn/hzjl/gjjldt/201012/t20101224\\\_3049390.html}, which is $\eta_{\rm mb}=0.40$ to 0.48 (varying with the frequency and elevation). After correction for $\eta_{\rm mb}$, for S140, the intensity is $T{\rm _{mb}(^{12}CO)}=68.8$ K and $T{\rm _{mb}(^{13}CO)}=29.0$ K. For W51D, the values are $T{\rm _{mb}(^{12}CO)}=91.7$ K and $T{\rm _{mb}(^{13}CO)}=37.5$ K. By comparing the spectra of the calibrators at our observing period, we can obtain the correction factor $C=T{\rm _{mb,calibrator}}/T{\rm _{ant,calibrator}}$ and use it to correct the intensity of the target source, which is $T{\rm _{mb,target}}=C\cdot T{\rm _{ant,target}}$. For S140, its observed $T{\rm _{ant}(^{12}CO)}$ has a daily variation of $\pm7\%$ (as due to the elevation change from $45^\circ$ to $65^\circ$), which is adopted as the uncertainty level in the flux calibration. We note that the AOSs also covered the ${\rm ^{18}CO}~(1-0)$ line ($f=109.78$ GHz). However, spectra often have a low signal-to-noise ratio or fluctuated baseline and are therefore not adopted for the subsequent analysis. 

The position-switch mode was used for the observation. The reference positions were examined to have no detectable emission feature within the observed bandwidth. Otherwise, if there is an emission feature, it should be separated from the line profile of the target. For the mapped regions, the mapping steps along the right ascension (RA) and declination (Dec.) directions are both 1 arcmin. The integration time was 1 minute for each point. The typical angular extent required to map the $^{13}$CO emission region is $7\times7$ arcmin$^2$. 

Another fraction of the observational data was obtained from the public 13.7-m telescope data base\footnote{http://www.radioast.csdb.cn/}. These data originally came from the observations of the star-forming regions selected from other tracers and Galactic CO line surveys. The $^{12}$CO and $^{13}$CO $(1-0)$ emission within 1.0-arcmin vicinity around the maser sources would be adopted for our study. With the archive data, we have found CO spectra for 160 methanol masers whose Galactic longitudes range from $l=9.6^{\circ}$ to $232.6^{\circ}$. All the data were calibrated using the same procedure as that performed for our own observation. The GILDAS software package\footnote{http://www.iram.fr/IRAMFR/GILDAS} was used for data reduction and plot. The program GraphPad Prism\footnote{GraphPad Prism version 5.00 for Windows, GraphPad Software, La Jolla California USA, www.graphpad.com} was also used for some statistical test and plot. The observed line parameters are shown in Table S1 in the online material, and the derived physical parameters are shown in Table 1.



\section{Results}
\subsection{Selection for the valid CO line components}
All the maser sources in our sample are detected in ${\rm ^{13}CO}~(1-0)$ emission with the 13.7-m telescope. It should still be further examined whether the $^{13}$CO gas is physically associated with the maser or is merely close in projection. This can be done by comparing the systemic radial velocity between the CO-line profile and the maser emission. In our sample, the velocity difference between the $^{13}$CO line and the maser is mainly concentrated in a range from 0 to 5 km s$^{-1}$, with a small fraction of the sources extending to higher values. 

If the maser source and the CO line have a small velocity difference, they should be spatially associated, while larger discrepancies may indicate them to be separated. A threshold velocity difference is needed to distinguish between the two cases. Using the PMO 13.7-m telescope, \citet{guan08} measured the $^{13}$CO velocity field in a number of molecular clouds and found a spatial variation from 2-5 km s$^{-1}$. Taking this result into account, we adopted a selection criteria including three cases: 

(1) if $|V_{\rm sys,maser}-V_{\rm sys,^{13}CO}|\leqslant 5$ km s$^{-1}$, the $^{13}$CO emission would be assumed to be associated with the maser source; 

(2) if $5<|V_{\rm sys,maser}-V_{\rm sys,^{13}CO}|\leqslant 10$ km s$^{-1}$, an association is still assumed but labeled as questionable ("q") in Table 1; 

(3) if $|V_{\rm sys,maser}-V_{\rm sys,^{13}CO}|>10$ km s$^{-1}$ the CO data would be excluded from our sample. 

Based on this selection criteria, 26 spectra lines have been excluded. We note that this criteria may not be fully adequate or complete. First, even if the CO lines and the maser source have similar velocities, the projection effect still exists. Second, another fraction of sources with a large velocity difference may be mistakenly excluded. After the velocity-based selection, 26 $^{13}$CO components have been excluded, which take up a relatively small fraction of the total number (203 components). Moreover, methanol masers are exclusively inhabited in the young high-mass star-forming sites which are abundant in molecular species. In this case, if a CO component coincides with it in both position and velocity, it should be less likely for them to be separated. A more accurate selection criteria is still to be considered. This is beyond the scope of the current work. 

Taking the multiple components into account, we obtained 177 validated $^{13}$CO gas components in 160 methanol maser sources in our sample. In the same Galactic longitude range there are 259 masers altogether in \citet{pestalozzi05}'s catalogue. In some regions, there are multiple maser spots located in a close vicinity (within 1 arcmin) with similar velocities. If these nearby maser spots are regarded as one single source, the total number of the maser sources would be 205. Then our spectra have covered 79 percent of these sources. The coordinates and systemic velocity of all the sources and components are presented in Table 1. In Figure 1, we present the source distribution in each Galactic longitude interval for the maser sources and the fraction included in our sample. As a major insufficiency, 38 maser sources in the Galactic range from $l=10^{\circ}$ to $50^{\circ}$ (26 percent of all) are not covered in our observation. In the catalogue of \citet{pestalozzi05}, the only parameter that can characterize the 6.7 GHz maser properties is their emission intensities. In the overlaid panel in Figure 1, we show the intensity distributions of the sampled and missed maser sources in this Galactic range. The two groups exhibit similar ranges and average levels for the maser intensities. The "missed" group tends to have slightly lower intensities than the "sampled" group. This indicates that our sample has a slightly better coverage for the strong masers. This is unlikely to significantly affect the sampling unless those missed ones have particularly special properties among the entire sample. In general, we suggest that our data should have an unbiased sampling for the 6.7 GHz maser sources in the corresponding Galactic longitude range.  

\subsection{Calculation of the physical parameters}
We derived the excitation temperature from the $^{12}$CO $(1-0)$ lines by assuming a high optical depth and a local thermal equilibrium (LTE), using 

\begin{equation}
T_{\rm ex}=\frac{h\nu}{k}[\ln(1+\frac{h\nu/k}{T_{\rm MB}({\rm ^{12}CO})/\eta_{\rm bf}+J_{\nu}(T_{\rm bg})})]^{-1}
\end{equation}

where $T_{\rm MB}({\rm ^{12}CO})$ is the main-beam brightness temperature of the ${\rm ^{12}CO}~(1-0)$, $T_{\rm bg}=2.73$ K is the brightness temperature of the cosmic microwave background. $J_{\nu}(T)=(h\nu/k)/[\exp(h\nu/kT)-1]$, and $\eta_{\rm bf}$ is the beaming filling factor. The variable $k$ is the Boltzmann constant, and $h$ is the Plank constant. As shown in the mapped regions (Figure S1 in online material) in most cases, the $^{13}$CO emissions extend to several arc minutes (Figure S1). In this case, $\eta_{\rm bf}\simeq1$ should be a reasonable assumption.

The $^{13}$CO optical depth is derived using~\citep{garden91}

\begin{equation}
\frac{T_{\rm MB}({\rm ^{12}CO)}}{T_{\rm MB}({\rm ^{13}CO})}\approx\frac{1-\exp({\tau_{12}})}{1-\exp({\tau_{13}})},
\end{equation}

wherein we assumed an optical depth ratio between ${\rm ^{12}CO}$ and ${\rm ^{13}CO}$ equal to their average Galactic abundance ratio, which is $\tau_{12}/\tau_{13}=[{\rm ^{13}CO}]/[{\rm ^{12}CO}]=50$. The ${\rm ^{13}CO}$ column density is calculated from 

\begin{eqnarray}
N({\rm ^{13}CO})=\frac{3k}{8\pi^3B\mu^2} \frac{\exp[hBJ(J+1)/kT_{\rm ex}]}{J+1}\times\nonumber \\ \frac{T_{\rm ex}+hB/3k}{1-\exp(-h\nu/kT_{\rm ex})}\int \tau_v \, dv,
\end{eqnarray}

where $B=54.89$ GHz is the rotation constant of ${\rm ^{13}CO}$, $\mu=0.11079$ Debye is the permanent dipole moment and $J=0$ is the rotational quantum number of the lower state. The ${\rm H_2}$ column density is calculated from $N({\rm H_2})={\rm [H_2]}/{\rm [^{13}CO]}\times N({\rm ^{13}CO})$. In the calculation, we assumed a ${\rm ^{13}CO}$ abundance of ${\rm [^{13}CO]/[H_2]}=1.0\times10^{-6}$, which is also adopted by the \citet{du08} for the Galactic IRDC survey using the 13.7-m telescope. The error in $N({\rm H_2})$ is estimated from the errors in $T{\rm _{mb}(^{12}CO)}$ and $T{\rm _{mb}(^{13}CO)}$ (both $\pm7\%$). The calculated $T_{\rm ex}$ and $N({\rm H_2})$ are presented in Table 1.

\subsection{Determination of the distances}
The distances to the sources are important for revealing their Galactic distribution and examining whether the measurement of the physical parameters is biased due the distance effect. For each source in our sample, we have searched in literature for the distance measurement. As a result, a few individual sources have a direct distance measurement from interstellar extinction, trigonometric parallax, or the luminosity of the exciting stars in the nearby (UC) H{\sc ii} regions. For a major fraction of the maser sources, however, we have to rely on the Galactic rotation curve to derive the kinematic distances $D_{\rm kin}$. The parameter $D_{\rm kin}$ is calculated from the Galactic rotation curve of \citet{brand93} and using the newly measured solar distance $R_0=8.3\pm0.23$ kpc, and the velocity $\Theta_0=239\pm7$ km s$^{-1}$ \citep{brunthaler11}. For all the valid $^{13}$CO components (including the "questionable" ones), the velocity difference between the maser and $^{13}$CO line and between the different $^{13}$CO line components are all assumed to reflect the variation in the same region. Consequently, an average radial velocity from the methanol masers and all the $^{13}$CO components is adopted to calculate $D_{\rm kin}$. In calculation, most sources in the first Galactic quadrant are subjected to the distance ambiguity, where there are two possible solutions for a given radial velocity (commonly known as near and far distances). There are several properties which can assist to resolve the ambiguity, including

(1) If the CO emission is associated with a (UC) H{\sc ii} region, which is observed in 21-cm HI absorption, and if the HI absorption has a systemic velocity larger than the maser source, the maser sources should be located at the far distance \citep{pandian08,anderson09,urquhart11}. 

(2) If the CO emission is associated with the 21-cm HI self absorption in both position and velocity, it should be at a near distance. In contrast, if the CO emission is associated with the HI emission, far distance would be adopted \citep[specified in][Figure 1 to 3 therein]{roman09}.  

(3) If the maser source overlaps with an infrared absorption feature, it is then assumed to be at the near distance, considering that there is a small possibility to see the CO gas, which is not associated with the IR absorption but located further behind.

(4) If the three properties above are unavailable, the one distance value closer to the nearby spiral arm of the two would be adopted. This strategy was used in the previous studies, including \citet{schwartz88} and \citet{faundez04}. For example, G60.56-0.17 has two distance values of $D_{\rm kin}=0.3$ and 8.0 kpc. Its near distance is even shorter than the Orion molecular cloud which is known to be the nearest high-mass, star-forming site \citep[414 pc,][]{menten07}. Instead, if we assume the far distance (8.0 kpc), it would be well located on the Perseus Arm.  


The distance determination and the references for all the sources are specified in Table 1. The major uncertainty in $D_{\rm kin}$ and $R_{\rm gc}$ is from the uncertainty in the fitted Galactic rotation curve \citep{brand93}. From the Figure 4b in \citet{brand93}, we can see that the rotation velocity has a deviation from the fitted curve with a scale of $\pm20$ km s$^{-1}$ for $R_{\rm gc}<8.5$ kpc. Based on this feature, we estimate the error by varying the Galactic rotation curve with this value of velocity. For $R_{\rm gc}>8.5$ kpc, the rotation curve begins to have large uncertainties ($>100$ km s$^{-1}$). To give a more reasonable estimate, we assume that the distance error of a source should be less than $1/2$ times the interval between two nearby spiral arms along its Galactic longitude, as measured from Figure 2.

\subsection{Distance-related errors in the physical parameters}
The large variation in our source distance may potentially influence our sampling and the measurement of the physical parameters. Towards longer distances, the beam area covers a larger spatial region, leading to two main effects: (1) the emission from a region smaller than the beam size would be weakened due to the beam dilution; (2) the different gas components with the beam area would all be averaged together. The second effect may cause $\Delta V({\rm ^{13}CO})$ to be overestimated at larger distances. To investigate the significance of the "distance effects", we plot the variation of the physical parameters including $T_{\rm ex}$, ${\rm N(H_2)}$ and $\Delta V({\rm ^{13}CO})$, as a function of the Heliocentric distance in Figure 3. We performed a linear least-square fit for each parameter. As a result, the variation of the physical parameters as a function of the distance is insignificant with a correlation factor of $<0.2$.  

As another examination, which can somewhat eliminate the errors in the distance measurement, we divided the data points into each 2-kpc interval from short to long distances and compare the statistical distributions in different intervals. The results are shown in the right panels in Figure 3. For each parameter, the data points in different intervals appear to be similar with each other. As quantitative examination, we carry out both the Student's t-test and the Mann-Whitney (M-W) U test. Both tests provide an estimate of whether one data set tends to have larger or smaller values than another one. The difference is that the t-test assumes a Gaussian distribution for the data set, while the M-W test assumes a random distribution. In Figure 3, we show the test results ($p$-values for the t-test and the M-W test) between some intervals, which are typical to illustrate the variation trend. The major results for the three parameters are as following:

(1) For the $T_{\rm ex}$ distribution, the 10-12 kpc interval shows significantly smaller $T_{\rm ex}$ compared to the other intervals. For example, it is different from the distribution in the 0-2 kpc interval with p=0.004 and 0.003 for the two tests. This suggests that the beam dilution may have weakened the distant sources. All the other intervals show similar $T_{\rm ex}$ distributions in both M-W and t tests ($p>0.05$) with none of them having significantly higher or lower values than the other intervals. This indicates that the beam dilution, if exists, may have only slightly influenced the $T_{\rm ex}$ measurement towards the most distant sources. 

(2) The $N({\rm H_2})$ (${\rm H_2}$ column density) distribution shows a similar result with that for $T_{\rm ex}$, that is, the 10-12 kpc interval tends to have smaller ${\rm N(H_2)}$ values than the other intervals (with $p=0.001$ and 0.005). This may also be explained as the beam weakening to the distant sources. 

(3) For the $\Delta V({\rm ^{13}CO})$ distribution, the 4-6 and 6-8 kpc intervals show higher $\Delta V({\rm ^{13}CO})$ than the other intervals, as indicated by the both tests. The other intervals have similar $\Delta V({\rm ^{13}CO})$ distributions with each other. From a theoretical point of view, if $\Delta V({\rm ^{13}CO})$ is subjected to the distance effect, the observed $\Delta V({\rm ^{13}CO})$ should roughly increase linearly at longer distances, because the beam area would statistically cover more gas components at larger distances. The observed trend is different. It shows a decreased $\Delta V({\rm ^{13}CO})$ in 8-12 kpc compared to the 4-6 and 6-8 kpc intervals. From Figure 2, we actually found that a large fraction of the sources in this distance range are located around the 5-kpc molecular ring and the edge of the central Galactic bar. These regions may actually have increased gas turbulence.

We note that the resolved distances are still suggestive more than definitive. If the ambiguity still exists, the distance effect can be evaluated at two extreme cases, which assumes that all the sources have near and far distances, respectively. The first case shows a very small difference compared to Figure 3 mainly because the near distances are already adopted in most cases. In the second case, which totally adopts the far distances, the resultant parameter distributions are shown in Figure S4 online material. The $T_{\rm ex}$ distribution appears to have a decline from $D=0$ to 10 kpc. The trend is more clearly shown on the statistical bars (right panel). However, the values rise again from $D=10$ to 18 Kpc, so that $T_{\rm ex}$ tends to have the smallest values in 8-10 kpc interval. A similar trend is also seen in the ${\rm N(H_2)}$ distribution (middle-left and -right panels). The figure shows that the 8-10 kpc interval has the lowest ${\rm N(H_2)}$ on average, as well as the largest distribution range. This feature is inconsistent with the expected beam dilution at larger distances which is expected to cause a linear decrease in $T_{\rm ex}$ and ${\rm N(H_2)}$. Instead, it may reflect a situation that many sources are mistakenly put into the 8-12 kpc interval. In the $\Delta V({\rm ^{13}CO})$ distribution (bottom panels), there is a slight increase for the average level with the distance similar as shown in Figure 3. In the mean time, 8-12 kpc intervals also show an abnormally crowded population compared to the other intervals, suggesting that the distances may be unreasonably determined.          

As a summary for the analysis above, Figure 3 shows that the beam dilution might slightly weaken the intensities for the sources at large distances, and the beam convolution might somewhat blend the different velocity components, but these effects are still weak. It is unlikely that the physical parameters have a noticeable systemic bias as due to the distance effect.

\subsection{Emission features in the mapped regions}
\subsubsection{Identification of the molecular cores}
The $^{13}$CO $(1-0)$ maps of all the observed regions are presented in Figure S1 in the online material. From the 33 mapped regions, we have identified 51 molecular cores altogether which are associated with or close to the maser sites within several arc minutes. Here, we define a $^{13}$CO core, as the gas condensation enclosed by the 50-percent contour in the $^{13}$CO $(1-0)$ integrated intensity map. In most regions the molecular cores can be well identified from the $^{13}$CO contours. There are a few sources with extended emission morphologies or overlapped multiple sources. In those regions we use ellipses to encircle the plausible $^{13}$CO cores. The ellipse is drawn to overlap with the part of the 50 \% contour that is not contaminated by the nearby cores or diffuse emissions. The sources with contamination include G14.33-0.64, G29.86-0.05, G30.82-0.05, G35.20-0.74, G35.20-1.74, G80.85+0.43, and G136.84+1.12. 

We measured the effective radius of the core using $\pi R_{\rm core}^2=A_{\rm core}$, where $A_{\rm core}$ is the core area encircled by the 50 \% intensity contour level in units of pc$^2$. We note that the core radius is subjected to a large uncertainty mainly because the spatial resolution largely differs with the distance. The values are therefore listed in Table S3 (online material) only for reference and not adopted for further quantitative analysis. The $T_{\rm ex}$ and $N({\rm H_2})$ of the molecular cores are estimated from the spectra at the position closest to the core center, and the values are also presented in Table S2.

\subsubsection{Infrared and radio continuum emissions}
To investigate the infrared (IR) emission of our molecular cores, we looked for the images from the database of GLIMPSE, MSX, and IRAS sky surveys\footnote{http://irsa.ipac.caltech.edu/}. In Figure S1, the $^{13}$CO $(1-0)$ contours are overlaid on the infrared image (gray scale). The IRAC 8 $\micron$ images from the GLIMPSE survey would be first considered if they are available. If a region has no GLIMPSE image, the MSX 8.28 $\micron$ data would be used instead. There is only one source (G213.70-12.6), which has neither MSX nor IRAC data. For this source, we adopted the IRAS 12 $\micron$ image to show the IR emissions. The IR emissions of our sources exhibit various morphologies. In some regions such as G10.62-0.38 and G12.79-0.19, the IR emissions clearly reveal multiple IR sources. In other regions, such as G15.03-0.68 and G30.82-0.05, the IR emissions are very extended and do not reveal any point-like sources. A major fraction of the IR-faint cores are actually not associated with the masers but the "additional" objects in the same region. Among the 33 cores, which coincide with the masers, there are only one source (G173.71) with no evident IR emissions. This suggests that the stellar heating is already significant and often generates extended IR emissions in most maser sources. 

Among the 30 $^{13}$CO cores with noticeable infrared emissions, G8.68-0.37 and G15.03-0.68 have IR emissions at the edge of the $^{13}$CO cores, and have much weaker IR emissions toward the core center. The observing center (zero offset) of each region is located at the mazer point which is frequently but not always coincident with the IR emission peak. In the regions of G29.86-0.05, G35.20-0.74, G045.44+0.07 and so on, the maser site does not have intense IR emission. Since the IR emission is necessary for the pumping the Class-II methanol masers~\citep{menten91}, there may still be stellar emissions in those cores, which are probably highly obscured by the dust.

To inspect the possible radio continuum emission in the molecular cores, we referred to the MAGPIS\footnote{http://third.ucllnl.org/gps/index.html, see also \citet{white05,helfand06}} 6 cm survey and other literature. The MAGPIS covers a Galactic longitude range of $350^{\circ}<l<49.5^{\circ}$ with a detection limit of $\simeq2.3$ mJy. There is another unbiased 6 cm survey performed by \citet{gregory91}, which covers the sky field between $0^{\circ}<{\rm Dec.}<75^{\circ}$ with a detection limit of $\sim5$ mJy. The combination of MAGPIS and Gregory's observation, have well covered our mapped regions. In addition, we also sought for other radio continuum observations for our fields. If a radio continuum source is located within the 50\% contour of a $^{13}$CO core, it is then regarded to be associated with the core. The pointing accuracy of all the free-free emission surveys is better than 10 arcsec. The radio continuum fluxes for the $^{13}$CO cores are listed in Table S2.

In our mapped region there are 17 cores which are likely associated with radio continuum emissions. In these regions, the radio emissions all coincide well with the IR emission, indicating that it should originate from the ionized gas near the young massive stars. The radio continuum intensities are presented in Table S2. If a core does not have radio emission as shown in MAGPIS, \citet{gregory91}, or other published observational studies, it would be regarded to be radio-faint. We note that the detection of the radio continuum highly depends on the observational sensitivity. For instance, G188.79+1.02, G188.95+0.89, G189.03+0.76, and G232.62+0.99 are below the detection limit of \citet{gregory91}. However, they actually have faint emission around 1 mJy (see Table S3). However, due to their weakness, we assume these cores to be radio-faint. The locations of radio continuum sources are labeled with open squares in Figure S1. With the radio-bright sources excluded, there are 18 masers sources, or 54\% that are left, which are likely pre-UC H{\sc ii} due to weak or undetected radio emissions.

\subsection{Infall and outflow signatures}
As an indication of high-velocity gas, some cores show plausible non-Gaussian line wings in $^{12}$CO $(1-0)$ spectra. To further identify the potential outflow motion originating in the ${\rm ^{13}CO}$ cores, we made p-v plot across each core along both the RA and Dec directions. If the contours on the p-v plot shows a significant swell toward the blue- and/or redshifted velocities, it is then suggested to be an outflow candidate. Using this selection rule, high-velocity wings are found in six of the 51 cloud cores: G29.86-0.06(1), G35.20-0.74(1), G45.44+0.07, G59.78+0.06(1), G94.58-1.79, and G173.49+2.42. The P-V diagrams and spectra at the core center are shown together in Figure S2a. The outflow in G35.20-0.74 has been previously detected in CO $(2-1)$ \citep{wu05}. On the other hand, we did not observe evident line wings in some sources with outflows detected, including G8.68-0.37 \citep{longmore11}, G10.62-0.38 \citep{liuhy10}, G10.47+0.03 \citep{gibb04}, and G31.41+0.31 \citep{olmi99}. In these regions, the outflow may have much smaller spatial scales, as in G10.62-0.38 where the outflow shows up as compact gas condensations at 10-arcsec scales \citep{liuhy10}. These outflows should have been diluted or contaminated with the gas turbulence on our observational scale due to the dilution by the large beam size. In Figure S2b, we present four sources which have relatively well-shaped outflow lobes. In the other two outflow candidates, the outflow lobes are largely contaminated with the surrounding gas and cannot be clearly identified. Their maps are thus not shown in the figure.    

Since class-I methanol masers show a broad association with the outflow and shock emissions \citep[][and the references therein]{chen09}, we also examined the association between the class-I masers and the outflows. As a result, we found that two outflow candidates, G45.44 and G59.78, are indeed associated with the class-I masers. The non-detection of outflow in other class-I masers may also be tentatively attributed to the beam dilution and gas contamination. In the recent extensive investigation for class-I masers \citep{valtts07}, 25 percent of the class-I sources are found to have outflows. This detection rate is much higher than that shown in our sample. In our sample, the low outflow detection towards class-I sources (75 \% of all) should be due to the small-scale outflows being less well traced by the CO $(1-0)$ emission or diluted by our beam.       

Plausible infall signature, which is a $^{12}$CO profile with blueshifted absorption and a single peak $^{13}$CO line~\citep{zhou93}, are detected in four cores. We estimated their infall rate from the $^{12}$CO line profiles. We first use Equation (9) in \citet{myers96} to derive the typical infall velocity from the $^{12}$CO line profile and then use Equation (3) in \citet{kw07} to estimate the infall rate. In calculation, the typical infall radius $(R_{\rm gm})$ is adopted as the core radius ($R_{\rm core}$) measured from the $^{13}$CO emission region. The derived infall parameters are listed in Table S3. We note that the low resolution and insufficient sampling may cause great uncertainty in $R_{\rm core}$ and, hence, the infall rate. The infall rates are presented only as tentative results and comparison to subsequent studies. 

\citet{kw07} observed 23 massive star-forming regions in ${\rm HCO^+}~(4-3)$ and detected infall motion in nine objects in the sample. The derived infall rates are on average around $10^{-4}$ $M_{\odot}$ year$^{-1}$, which are much smaller than the tentative measurement here. In G8.68-0.37 where the infall signature is detected in both CO and ${\rm HCO^+}$, we estimated an infall velocity of 0.7 km s $^{-1}$, which appears consistent with their value, and our different infall rate is mainly due to the large core radius estimated from the low-resolution $^{13}$CO map. Except for G34.26, all the infall candidates in \citet{kw07}, including G8.68, G10.47, G15.03, G109.86, G133.94, and G192.6 are also in our sample. Besides G8.68-0.37, our CO spectra do not show evidence of infall signatures in any of the other sources. The object G35.20-0.74 was detected with a blue profile in CO $(2-1)$ line \citep{wu05}, which is not evident in the current CO $(1-0)$ line. The possible reason is that the two molecular species are tracing different gas components. The CO, which mainly traces the low-density gas, may have a flatter temperature profile, so that self-absorption is not significant. The difference between the CO and ${\rm HCO^+}$ may imply an inside-out mode for the core collapse in these regions, which can be tested by future high-resolution studies.

\subsection{Comments on individual objects}
The properties of the sources noteworthy are described as follows:

1. G8.68-0.36

This region contains a single compact $^{13}$CO core. The UC H{\sc ii} region G8.68-0.36~\citep{wood89} is associated with the strong IR emission in the IRAC 8 $\micron$ bands. The UC H{\sc ii} region is located west to the $^{13}$CO core center. Recent high-resolution observation revealed a strong bipolar outflow originated from the core~\citep{longmore11}.

2. G10.47+0.03

This region has a single compact $^{13}$CO core. The core is elongated from the southeast to northwest. The densest part of the molecular core is associated with an extended 8 $\micron$ emission, which shows a clustered environment at the center of the molecular core. The high-resolution study~\citep{pascucci04} reveals a group of young high-mass stellar objects at different evolutionary stages.

3. G10.62-0.38

This field contains a roughly spherical $^{13}$CO core. The IRAC 8 $\micron$ image shows several stellar sources concentrated at the center and diffuse emission stretching to the outer part, and a cluster is being formed at the center. Ionized accretion flow was detected in the UC H{\sc ii} region~\citep{keto02,sollins05}. High-resolution study revealed a complicated outflow system arising from the stellar objects at the center~\citep{liuhy10}.

4. G12.79-0.19 

This region shows a compact $^{13}$CO core, which is slightly elongated from northeast to southwest. The $^{12}$CO $(1-0)$ line at the center shows a blue profile, which indicates the infall motion (Figure 3). It also has a broad emission feature extended to the redshifted side of the line profile ($V_{\rm lsr}=40$ to 50 km s$^{-1}$). However, this emission feature also show up in the $^{13}$CO $(1-0)$, indicating a considerable optical depth. It is therefore more likely to be a different velocity component rather than merely the outflow. In the IRAC 8.0 $\micron$ image, we can see three bright stellar sources at the core center with one of them coincident with the radio continuum source. In their vicinity, there are several extended filamental structures which extend further away from the center. They can be hot gas and dust expanding from the center, or the ambient gas compressed and swept by the stellar emission.   

5. G14.33-0.64 

This region shows extended CO emission. The $^{13}$CO emission is resolved into four major dense cores. Core-1 and core-2 are not fully separated from each other, but their IR sources are well separated. Thus, we expect them to be different molecular cores. Core-3 and 4 are free of IR emissions. The MAGPIS 6 cm survey shows no detectable radio continuum emission in this (entire) region.

6. G29.86-0.05

In this region, the bulk of the CO emission is located to the north of the center (i.e., the CH$_3$OH maser location). Therefore the methanol maser may arise from a less extended core, which is currently unresolved. The emission region is extended and resolved into four major dense cores. Core-1 is associated with a bright IRAC point source. It also drives an outflow, as shown by the broad line wings (Figure S1). The remaining three cores are much fainter in the CO $(1-0)$ and IR emissions.

7. G30.82-0.05 

This region exhibits an extended CO emission feature. We have identified three $^{13}$CO cores corresponding the strongest emission peaks. We examined the C$^{18}$O emission and confirmed the location of the three strongest cores. The cores all reside at the edge of the IR emission region. The methanol maser is associated with core-1. Core-2 and core-3 have no methanol masers but are detected in the radio continuum emission.

8. G31.41+0.31 

This source is a frequently studied high-mass star-forming region. In our observation, it shows a quite extended and irregular emission region with a scale of $\sim6'$ (9 pc). It may therefore contain many unresolved sub-structures. Nevertheless, the $^{13}$CO emission peak coincides well with the methanol maser and the hot molecular core observed at higher resolutions \citep[][and reference therein.]{girart09}. This core has a CH$_3$CN rotational temperature of $\sim164$ K~\citep{beltran05}, which is much higher than the $^{12}$CO excitation temperature (17.5 K). The difference should be mainly because the CO $(1-0)$ intrinsically traces colder gas components than the higher-excited lines and the high-density molecular tracers. This should be universal from the highest $T_{\rm ex}$, the comparison between the sources at different evolutionary stages suggests that the stellar heating can still noticeably affect the CO temperature (see Section 4.2).

9. G35.20-1.74

In the MSX 8 $\micron$ image, this region shows a prominent contrast between the bright IR emission in the east and the broad dark region in the west. The $^{13}$CO $(1-0)$ line reveals two dense cores near the IR emission. There is also a large area of extended CO emission in the west, which is coincident with the IR-dark region, wherein we do not identify any well-shaped cores. There is a strong radio continuum emission \citep{gregory91}, as marked with the red square. Since the radio continuum sources are located at a position, where the two cores are overlapped, we assume that the both cores are associated with the radio emission. Core-1 exhibits a self-absorption in $^{12}$CO $(1-0)$, which may indicate an infall motion.

10. G080.85+0.43

There are four dense cores in an alignment from northwest to southeast. Two of them are associated with the MSX point sources. Core-1 is detected in radio-continuum emission. The other cores are radio-quiet.

11. G183.34+0.59

This region shows a relatively compact core which is associated with a weak MSX point source. A strong MSX source is located on the western edge of the core, which also shows bright radio continuum emission. The $^{13}$CO emission shows a sharp decline toward the IR source, indicating that the molecular gas is being destroyed in the UC H{\sc ii} region.

\section{Discussion}
\subsection{Variation with the Galactocentric distance}
In Figure 4 we plot the variation of $\Delta V({\rm ^{13}CO})$, $N({\rm H_2})$, and $T{\rm _{ex}(^{12}CO)}$ as a function of the Galactocentric distance $R_{\rm gc}$ for our maser sources. We note that $R_{\rm gc}$ always has a single solution and is unaffected by the distance ambiguity. As a major result, the line width shows a noticeable decrease with $R_{\rm gc}$. At $R_{\rm gc}=4$ kpc, the average line width is $\Delta V=8$ km s$^{-1}$, and it drops to $\Delta V=2.5$ km s$^{-1}$ at $R_{\rm gc}=9$ kpc. The two statistical tests show that the different intervals vary with each other with $p<0.05$ in most cases. The $\Delta V-R_{\rm gc}$ relation may suggest that the gas becomes more turbulent in the maser sources closer to the Galactic center. The inner Galaxy on average has a higher star-and-gas density than the outer region. In a denser environment, the molecular cores can become more turbulent when affected by two aspects: a more frequent dynamical interaction with nearby molecular cores or clouds and the energy input from the stellar wind and radiation. Besides, the line broadening due to the optical depth should also be considered. Assuming that the optical depth is a Gaussian-shaped function of the radial velocity \citep{myers96}, $\tau(V)=\tau_0\exp[-(V-V_{\rm sys})^2/2\sigma^2]$, where $\tau_0$ is the optical depth at the systemic velocity $V_{\rm sys}$ and $\sigma$ is the velocity dispersion along the line of sight and can be measured from the intrinsic line width using $\sigma=\Delta V/\sqrt{8\ln2}$. The observed line profile would be $T_{\rm mb}(V)=J_\nu(T_{\rm ex})[1-\exp(-\tau(V)]$. From these equations, the ratio between the observed line width ($\Delta V_{\rm obs}$) and the intrinsic line width (broadening factor) is derived to be 

\begin{equation}
\frac{\Delta V_{\rm obs}}{\Delta V}=[\frac{\ln{\tau_0/\ln[2/(1+e^{-\tau_0})] }}{\ln 2}]^{1/2}.
\end{equation}

For all our ${\rm {13}CO}$ components, the broadening factor in $^{13}$CO $(1-0)$ turns out to be less than 1.5. Thus its contribution to $\Delta V_{13}$ and the $\Delta V_{13}$-$R_{\rm gc}$ correlation would not be significant, and we suggest that the cores at smaller $R_{\rm gc}$ may indeed have larger velocity dispersions. A similar trend between $\Delta V$ and $R_{\rm gc}$ was also suggested for the CO gas in the IRDCs \citep{du08}.

The other two parameters, $N({\rm H_2})$ and $T{\rm _{ex}(^{12}CO)}$ do not show significant Galactic variation. For the $N({\rm H_2})$ distribution, the 2-4 and 10-12 kpc intervals have very large probabilities to have similar distributions. The $T{\rm _{ex}}$ distribution shows a slight increase toward large $R_{\rm gc}$ with the difference reaching the significant level $(p>0.05)$ between 0-2 and 10-12 kpc intervals. However, its variation scale is still much weaker than the trend in $\Delta V$.

\subsection{Correlation between the physical parameters}
The relation between the different physical parameters are shown in Figure 5. Figure 5a shows that the $^{12}$CO excitation temperature significantly increases with the H$_2$ column density (with a slope of $3.3\pm0.3$ and a correlation factor of $R=0.4$). \citet{krumholz10} performed a hydrodynamic simulation about the molecular core collapse and star formation therein. The simulation shows that a higher surface density can significantly facilitate the formation of high-mass stars ($M>10~M_{\odot}$), thus lead to an increased temperature in the natal gas-and-dust core. In their densest case of $\Sigma=10$ g cm$^{-2}$ or $N({\rm H_2})=2\times10^{24}$ cm$^{-2}$, the final temperature in the surrounding gas exceeds $1000$ K as heated by the newly formed massive stars. In contrast, only a group of low-mass stars can be formed in the low-density case of $\Sigma=0.1$ g cm$^{-2}$, and the final temperature is $\sim100$ K. The simulation result shows a reasonable consistency with the observed $T_{\rm ex}-N({\rm H_2})$ correlation. Namely, the formation rate of the massive stars and/or heating efficiency would increase in higher-$N({\rm H_2})$ regions.

The above explanation have some apparent uncertainties. The major question is whether the internal stellar heating can noticeably change $T{\rm _{ex}(^{12}CO)}$. Due to the large optical depth in $^{12}$CO, its surface temperature might be insensitive to the internal stellar heating and instead, more affected by the external emissions. Based on our observational data, there is one way to examine this problem. Our mapped regions have provided a number of $^{13}$CO cores that are not associated with radio continuum and/or IR emissions. These cores can naturally serve as a control group for us to evaluate the significance of the internal heating. 

We first compared the temperatures in the radio-bright and radio-faint cores ($F_{\rm cm}<6$ mJy). Their distributions are shown in Figure 6a. The Figure shows that the radio-bright group are apparently inclined to high temperatures. The M-W and t tests both indicate that radio-bright cores have significantly higher $T{\rm _{ex}}$ than the radio-faint ones. The average temperatures of the two groups are 26.4 K and 20.5 K, respectively. The second comparison is based on their IR emissions, as shown in Figure 6b. The IR-bright and IR-faint groups have average temperatures of 24 K and 16 K with the contrast being slightly larger. This is within our expectation, since the IR-faint cores are actually a sub-sample within the radio-faint group. The comparison suggests that the cores without IR and radio continuum emissions indeed tend to have the lower $T{\rm _{ex}(^{12}CO)}$. 

In the case that $T{\rm _{ex}(^{12}CO)}$ is sensitive to the internal stellar heating, the observed $T_{\rm ex}-N({\rm H_2})$ correlation would favor the explanation that the stellar emission is actually more intense and/or more efficient in the higher-$N({\rm H_2})$ cores. Nevertheless, our current data is still insufficient to confirm the physical connection between $N({\rm H_2})$ and star-forming activities. For example, due to the low resolution, we cannot determine whether the IR and radio sources are embedded in or separated from the densest gas in a given core. It calls for denser molecular tracers to more accurately determine the temperature and $N({\rm H_2})$ distributions and to better reveal their correlation with the star-forming processes. 

The relation between $\Delta V({\rm ^{13}CO})$ and $N({\rm H_2})$ is investigated and shown in Figure 9. The value of $\Delta V({\rm ^{13}CO})$ appears to have no significant correlation with the ${\rm H_2}$ column density. A linear fit provides a slope of 0.28 and the correlation factor of $R=0.09$ which are both much smaller than the corresponding values in the $T_{\rm ex}-N({\rm H_2})$ relation. This may be explained as a result from the Larson relation \citep{larson81}. \citet{larson81} showed that, $\Delta V$ is sensitive only to the spatial size, which reflects the nature of the hierarchical turbulence, and in the mean time, the number density is anti-correlated with the spatial size. These two factors would cause the column density to be almost independent on the spatial size, which is consistent with the observed weak correlation between $\Delta V({\rm ^{13}CO})$ and $N({\rm H_2})$.  

\subsection{Comparison between the maser sources and the infrared-dark cores}
Besides the CH$_3$OH masers, the IRDCs also serve as a potential reservoir for high-mass cores at the earliest evolutionary stages \citep[e.g.][]{rathborne06}. \citet{du08} have performed a single-point and mapping survey of 61 IRDCs, selected from the prominent absorption feature at the MSX 8-micron band that uses the PMO 13.7 m telescope in the three isotopic CO $(1-0)$ lines. Our sample only has two sources (G35.19 and G173.71) in common with \citet{du08}. Even in these two regions, our $^{13}$CO cores have an offset larger than 5 arcmins from the observing center of \citet{du08}. A large fraction of the maser sources are also associated with IR absorption feature, as shown in \citet[][also see our Table 1]{peretto09}, but do not appear in their IRDC sample probably because the absorption is much less prominent than those shown on the MSX 8 $\micron$ band. 

In Figure 7, we plot the distributions of the physical parameters in the maser and IRDC samples. The two samples turn out to have largely different $N({\rm H_2})$ distributions. The IRDCs have a major fraction with $N({\rm H_2})>4\times10^{22}$ cm$^{-2}$ and have no distribution below $10^{22}$ cm$^{-2}$. The methanol masers, in contrast, are mainly concentrated at a small $N({\rm H_2})$ range, with the distribution largely decreasing at $N({\rm H_2})>10^{22}$ cm$^{-2}$. The IRDC sample is selected from the intense absorption features at MSX 8 $\micron$ band; thus, they are expected to have large $N({\rm H_2})$. Moreover, the difference between the two samples implies that the source selection based on the IR absorption feature could be biased toward high-$N({\rm H_2})$ regions, which misses out a considerable amount of high-mass star-forming cores inhabited in the clouds with moderate $N({\rm H_2})$.  

As shown in the middle and bottom panels in Figure 7, the two samples have similar $T_{\rm ex}({\rm ^{12}CO})$ and $\Delta V({\rm ^{13}CO})$ distributions much more similar, except that the maser sample has larger dispersion in $T_{\rm ex}({\rm ^{12}CO})$, which extends from 5 to 40 K. In comparison, the IRDCs show a concentration between 10 and 30 K. The possible explanation is that the methanol masers sources are generally at later evolutionary stages, thus having more various physical conditions. The stellar emission can either increase the temperature or dissipate the molecular gas and drive chemical reactions to consume CO. The latter aspect would cause a decrease in both the optical depth and beam filling factor, and lead to the  smaller values in the observed $T_{\rm ex}$ (5 to 10 K), as compared to the IRDCs. 

\section{Summary}
We performed an observational study with the $^{12}$CO and $^{13}$CO $(1-0)$ lines of 160 methanol masers and made mapping observation of a fraction of them (33 sources). We estimated the physical parameters and examined their Galactic variation and statistical properties. Our main results include four aspects:

(1) We examined physical parameters as a function of the Galactocentric distance ($R_{\rm gc}$). As a result, the $^{13}$CO line widths show evidence of a decrease from inner to outer Galaxy. This may reflect a feature that star-forming regions toward the inner Galaxy are more turbulent. The other two parameters, $T_{\rm ex}({\rm ^{12}CO})$ and $N({\rm H_2})$, do not show a prominent variation trend throughout the Galaxy.

(2) We examined the correlation between the different physical parameters. As a result, the ${\rm ^{12}CO}$ excitation temperatures show a noticeable increase with the ${\rm H_2}$ column densities. This provides consistency with the numerical simulation that suggests an increasing high-mass star formation and heating efficiency of higher-$N({\rm H_2})$ regions. This explanation is possible because the IR- and radio-bright cores show higher ${\rm ^{12}CO}$ excitation temperatures than those with an absence of heating sources. 

(3) We compared the physical parameters of the molecular cores in our maser sample to those of the infrared dark clouds. The maser sources tend to have noticeably lower $N_{\rm H_2}$ compared to the IRDCs. As an indication, the intense IR absorption features might miss out on some high-mass star-forming cores in low-$N({\rm H_2})$ regions. The other parameters show less significant distinctions between the two samples. 

(4) In our mapped regions, we have identified 51 cores from $^{13}$CO $(1-0)$ emission. Among those cores we found six outflow and four infall candidates. There are 17 cores associated with significant radio continuum emission and 42 cores with significant IR emissions. This indicates that the stellar emissions are already noticeably heating the surrounding gas in the methanol-maser sources, while the ionization is still at a less developed stage. 

\section*{Acknowledgment}
We are grateful to the working staff in the Qinghai station of the Purple Mountain Observatory 13.7 meter telescope for their valuable support to our observations, and the technical support from State Key Laboratory of Radio Astronomy. We are grateful to the anonymous reviewer for the constructive comments and suggestions. This work is supported by the NSFC grants of No.11373009 and the NKBRP grants of 2009CB24901 and 2012CB821800.

\bibliography{rwl2013_bib}
\clearpage

\onecolumn{

\begin{figure}
\centering
\includegraphics[angle=0,width=0.6\textwidth]{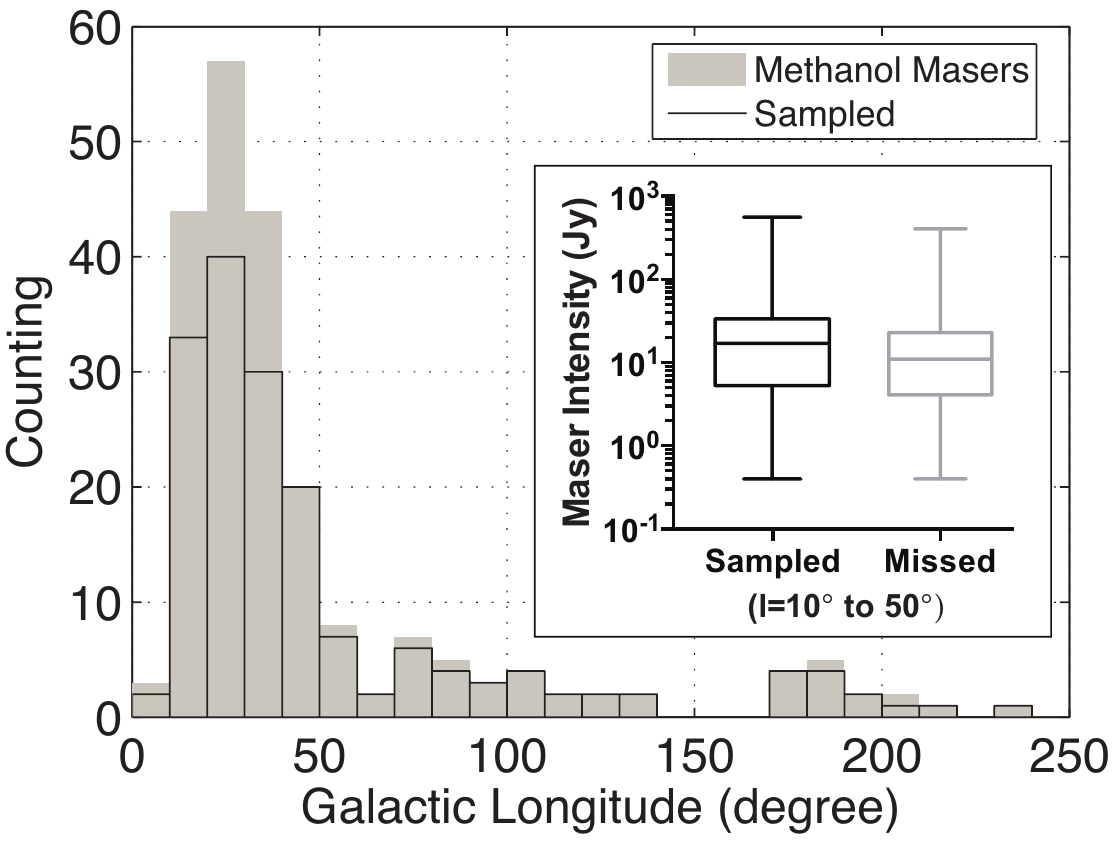} \\
\caption{The number of sources in each 10-degree Galactic longitude interval. The gray and black-line columns represent the full sample of the methanol masers \citep{pestalozzi05} and the fraction covered by our sample, respectively. {\bf Overlaid inner panel:} The distributions of the maser intensities for the sampled and missed sources in $l=10^\circ$ to $50^\circ$ range. For each statistical box, the entire error bar (whisker) represents the overall distribution, the box and its inner segment represent the 25\% to 75\% range and the median value, respectively.}
\end{figure}

\begin{figure}
\centering
\includegraphics[angle=0,width=0.5\textwidth]{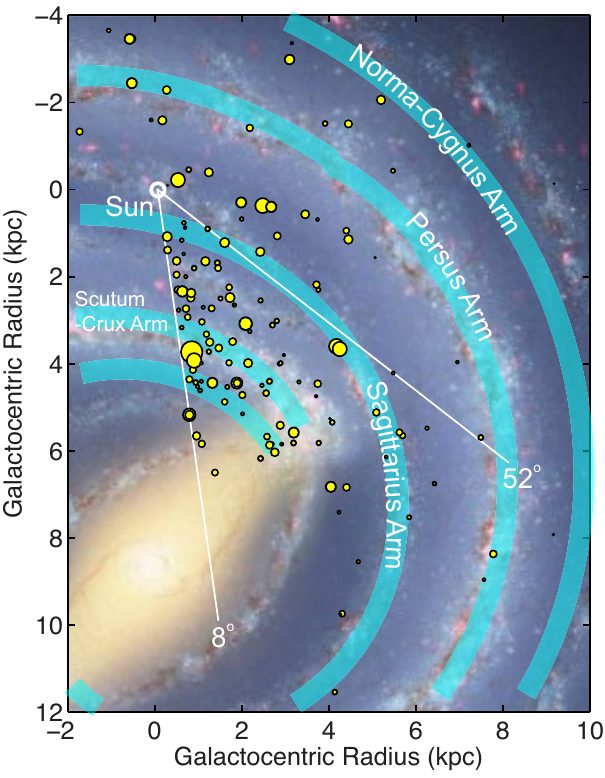} \\
\caption{The distribution of our observed sources on the Milky Way disk. The source sizes are proportional to their ${\rm ^{13}CO}$ line intensity. The thick faint blue lines represent the distribution of the spiral arms \citep{russeil03} in our observed field. The background image is the artist's view of the Galactic morphology (R. Hurt: NASA/JPL-Caltech/SSC, with a more detailed introduction in \citet{urquhart11}). }
\end{figure}

\begin{figure}
\centering
\includegraphics[angle=0,width=0.9\textwidth]{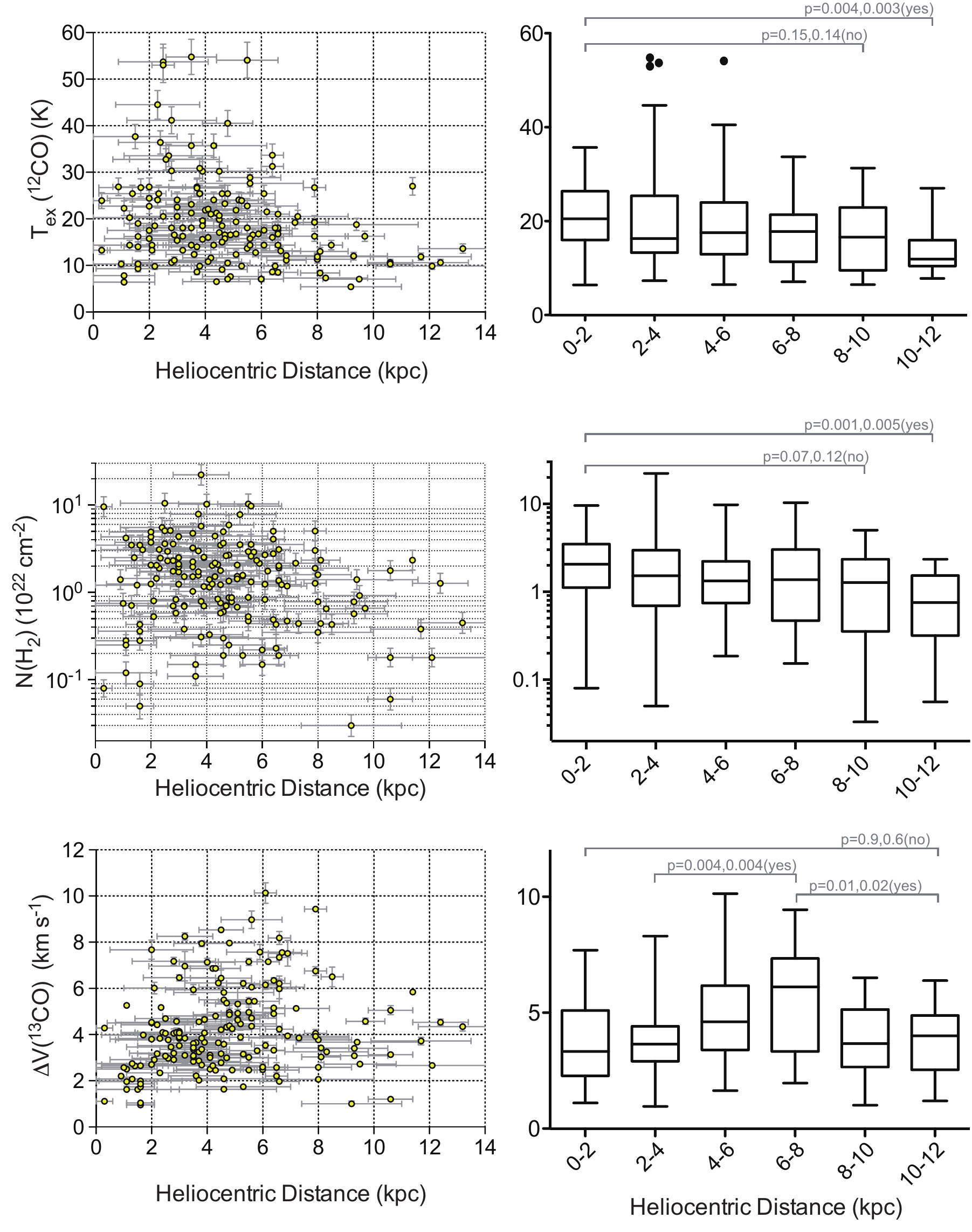} \\
\caption{The distribution of the physical parameters with the heliocentric distance. {\bf Left Panels:} All the data points. {\bf Right Panels:} Statistical box plots for the data distribution in each 2-kpc interval. The dots represent the outliers, which are apparently isolated from the major fraction of the data points. For each statistical box, the entire whisker represents the overall distribution. If there are isolated data points (based on eye selection), the whisker then represents the remaining data range. The box and its inner segment represent the 25\% to 75\% range and the median value, respectively. We selected interval pairs, which may typically reflect the variation trend. The interval pairs are indicated with gray segments above the whiskers. Labeled on the segment are the $p$ values from the t test and M-W test respectively, and the test result ("yes" or "no" in the brackets) for whether the two intervals have significantly different distributions.}
\end{figure}

\begin{figure}
\centering
\includegraphics[angle=0,width=0.9\textwidth]{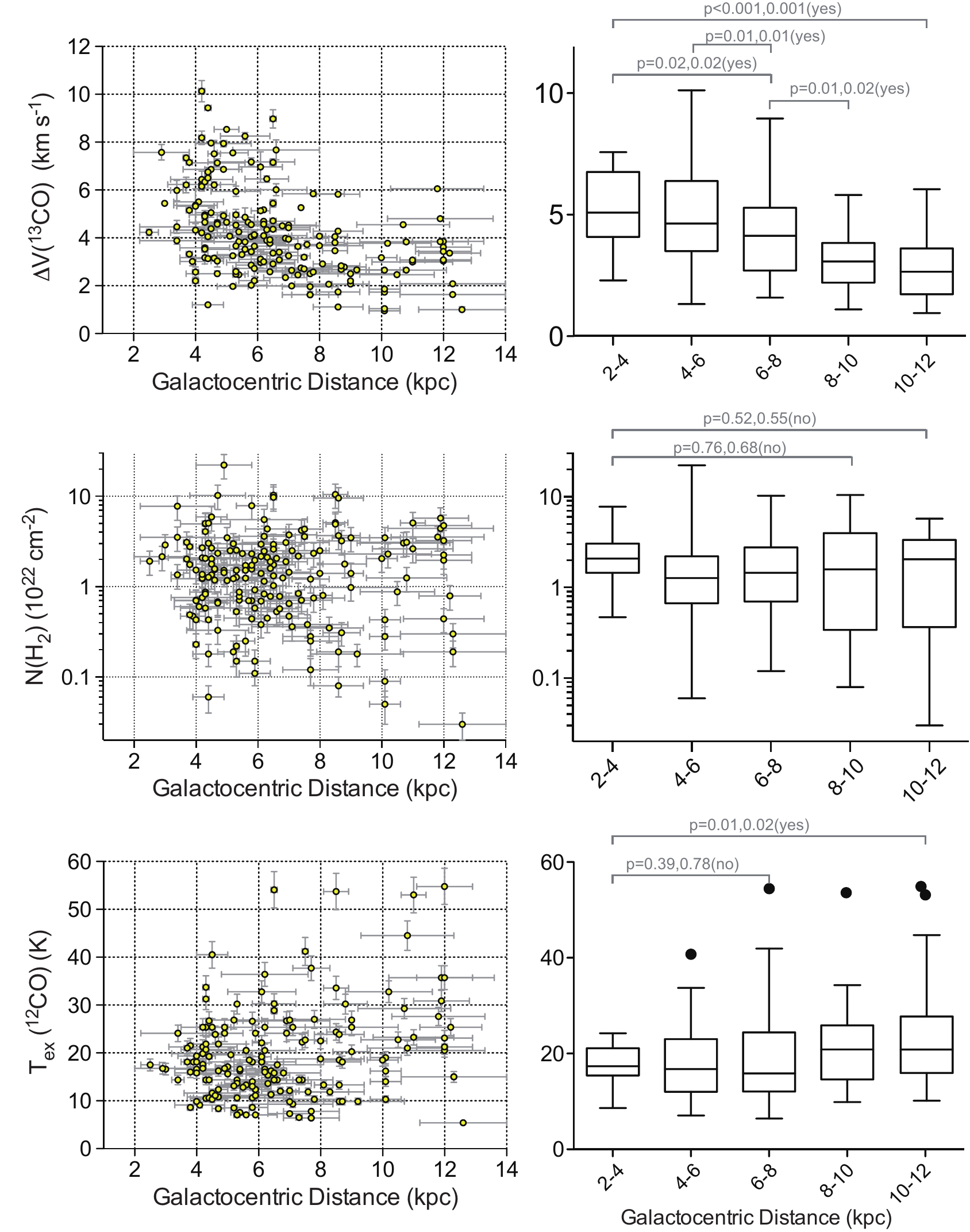} \\
\caption{The distribution of the physical parameters with the Galactocentric distance. The first panel shows the data points for the ${\rm ^{12}CO}$ excitation temperature. The other panels show the statistics of the parameters in each 2-kpc interval. For each statistical box, the entire whisker represents the overall distribution. If there are isolated data points (based on eye selection), the whisker then represents the remaining data range. The box and its inner segment represent the 25\% to 75\% range and the median value, respectively. We selected interval pairs which may typically reflect the variation trend. The interval pairs are indicated with gray segments above the whiskers. Labeled on the segment are the $p$ values from the t test and M-W test respectively, and the test result ("yes" or "no" in the brackets) for whether the two intervals have significantly different distributions.}
\end{figure}

\begin{figure}
\centering
\includegraphics[angle=0,width=0.9\textwidth]{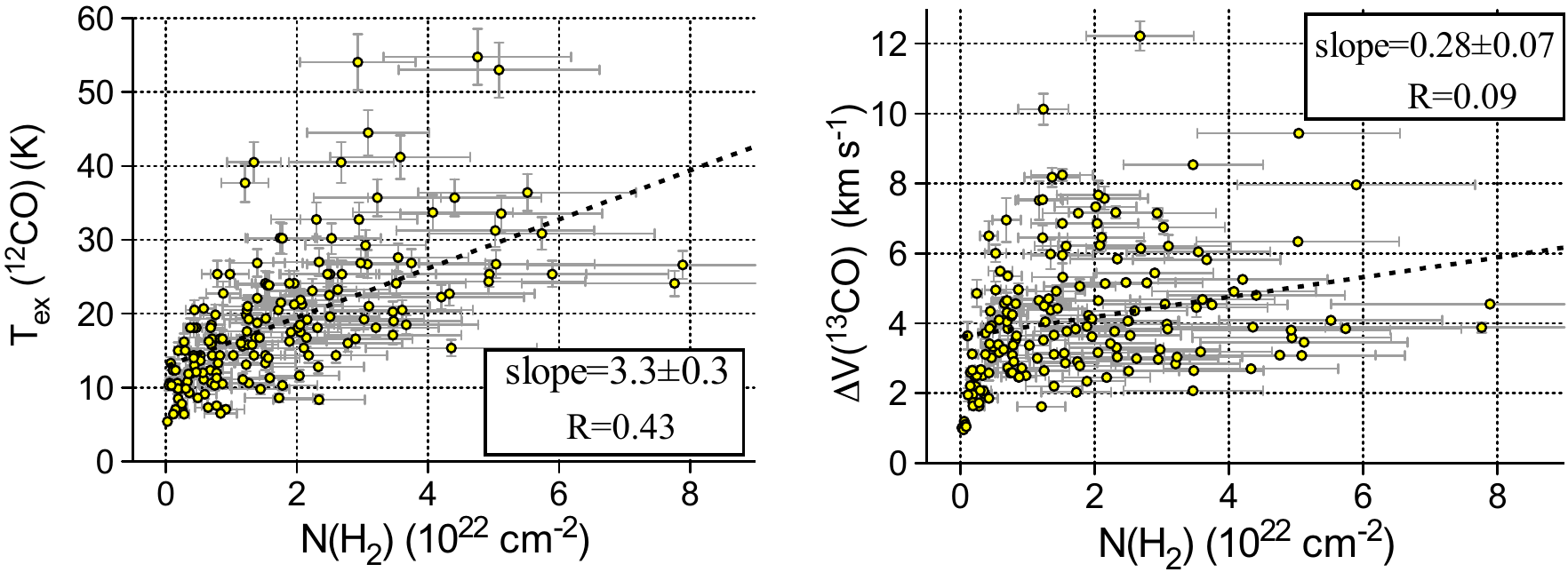} \\
\caption{The variation of the ${\rm ^{12}CO}$ excitation temperature and the ${\rm ^{13}CO}$ line width as a function the ${\rm H_2}$ column density. The dashed lines represent the best-fit linear correlation. }
\end{figure}

\begin{figure}
\centering
\includegraphics[angle=0,width=0.7\textwidth]{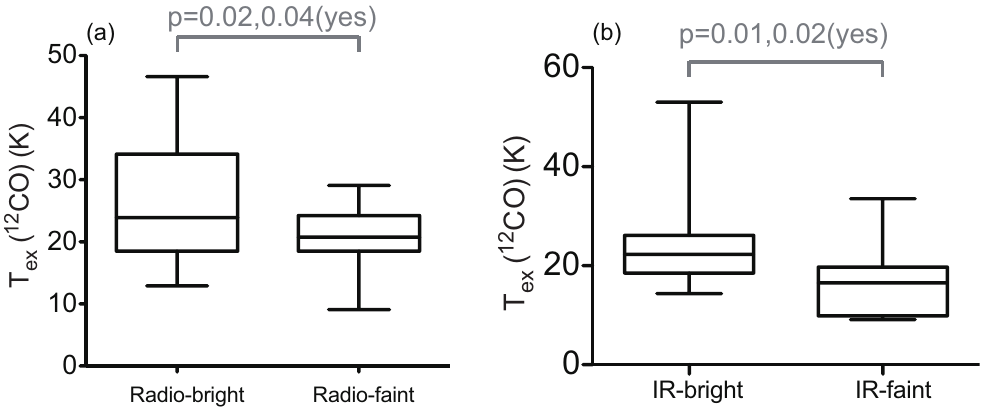} \\
\caption{The statistical comparison for the excitation temperatures between the different subsamples of the mapped cores. {\bf Left panel:} The radio-bright and radio-faint cores. {\bf Right panel:} The IR-bright and IR-faint cores. For each data set, the whisker represents the overall distribution; the box and inner segment represent the 25\% to 75\% range and the median value, respectively.}
\end{figure}

\begin{figure}
\centering
\includegraphics[angle=0,width=0.6\textwidth]{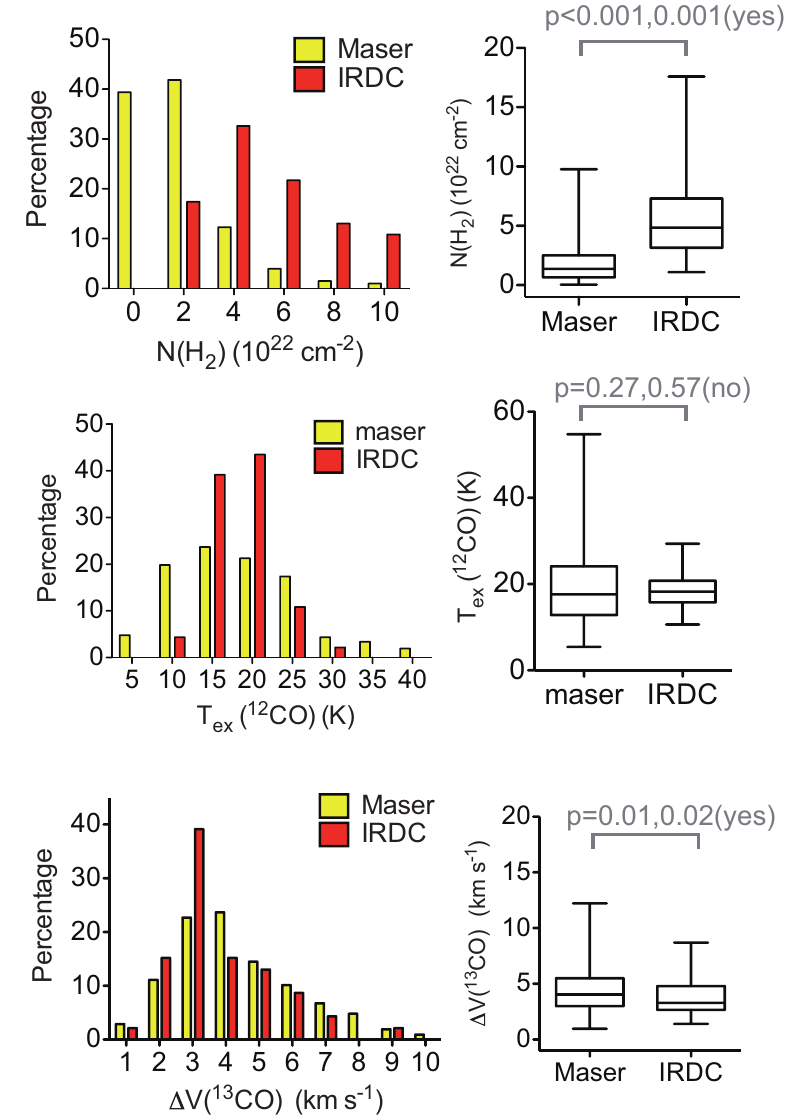} \\
\caption{Comparison of the physical parameters in our maser sources and the IRDCs \citep{du08}. The left panels show the parameter distribution in each interval. The right panels show the overall statistics. For each data set, the whisker represents the overall distribution; and the box and its inner segment represent the 25\% to 75\% range and the median value, respectively.}
\end{figure}


\begin{longtab}
\begin{landscape}
\small{
\begin{longtable}{@{\extracolsep{\fill}}lllllllllllll}
\caption{Physical parameters of the maser sources.} \\
\hline
Name                & $V_{\rm LSR}({\rm ^{13}CO})$   &  $V_{\rm LSR}{\rm (maser)}$   &  Relation    &  $D_{\rm kin}$          &  $D_{\rm other}$   &  $D_{\rm adopt}$         &  $R_{\rm gc}$     &  $\Delta V{\rm (^{13}CO})$   &  $T_{\rm ex}({\rm ^{12}CO})$    &  $N({\rm H_2})$           \\
\quad               & (km s$^{-1}$)                  &  (km s$^{-1}$)                &  \quad       &  (kpc)                  &  (kpc)             &  (kpc)                   &  (kpc)            &  km s$^{-1}$                 &  K                              &  $10^{22}$ cm$^{-3}$      \\
(1)                 & (2)                            &  (3)                          &  (4)         &  (5)                    &  (6)               &  (7)                     &  (8)              &  (9)                         & (10)                            &  (11)                    \\ 
\hline \endfirsthead
\caption{continued.}\\                                                                                                                                                                                                                                                                                                           
\hline                                                                                                                                                                                                                                                                                                                         
Name                & $V_{\rm LSR}({\rm ^{13}CO})$   &  $V_{\rm LSR}{\rm (maser)}$   &  Relation    &  $D_{\rm kin}$          &  $D_{\rm other}$   &  $D_{\rm adopt}$         &  $R_{\rm gc}$     &  $\Delta V{\rm (^{13}CO})$   &  $T_{\rm _{ex}}({\rm ^{12}CO})$    &  $N({\rm H_2})$           \\
\quad               & (km s$^{-1}$)                  &  (km s$^{-1}$)                &  \quad       &  (kpc)                  &  (kpc)             &  (kpc)                   &  (kpc)            &  km s$^{-1}$                 &  K                              &  $10^{22}$ cm$^{-3}$      \\
(1)                 & (2)                            &  (3)                          &  (4)         &  (5)                    &  (6)               &  (7)                     &  (8)              &  (9)                         & (10)                            &  (11)                    \\ 
\hline \endhead                                                                                                                                                                                           
008.68-0.36a        &  $042.857(0.110)$              &  $043.0$                      &    yes       &   5.2*,11.8[c1]         &  \quad             &  $05.2(1.3)$             &  $03.4(1.2)$      &   $03.88(0.15)$              &   $24.1(1.7)$                  &    $07.77(02.33)$ \\  
008.68-0.36b        &  $038.901(0.194)$              &  \quad                        &    yes       &   \quad                 &  \quad             &  $05.2(1.3)$             &  $03.4(1.2)$      &   $04.46(0.27)$              &   $24.1(1.7)$                  &    $03.52(01.06)$ \\
009.62+0.19         &  $003.611(0.039)$              &  $001.0$                      &    yes       &   0.9*,16.5[c1]         &  5.7[e8]           &  $05.7$                  &  $03.0$           &   $05.44(0.10)$              &   $16.6(1.2)$                  &    $02.90(00.87)$ \\
010.32-0.26         &  $038.979(0.076)$              &  $039.0$                      &    yes       &   4.4*,12.2[c1]         &  \quad             &  $04.4(1.2)$             &  $04.2(1.1)$      &   $06.24(0.19)$              &   $21.3(1.5)$                  &    $02.08(00.62)$ \\
010.47+0.02         &  $073.485(0.218)$              &  $073.0$                      &    yes       &   5.9*,10.8[c1]         &  \quad             &  $05.9(1.2)$             &  $02.9(0.9)$      &   $07.58(0.33)$              &   $16.8(1.2)$                  &    $02.15(00.65)$ \\
010.62-0.29         &  $-13.307(0.021)$              &  $-08.0$                      &    yes       &   18.9                  &  \quad             &  $18.9(1.5)$             &  $10.7(1.5)$      &   $04.55(0.04)$              &   $29.3(2.1)$                  &    $03.05(00.92)$ \\
\hline 
\end{longtable} 
\begin{flushleft}
An introduction for each column: \\
(1){ The source names, which are also their Galactic coordinates. If one source has multiple velocity components, letter "a, b, c..." are followed to label them.} \\
(2){ The systemic velocity of the ${\rm ^{13}CO}~(1-0)$ lines. } \\
(3){ The systemic velocity of the maser sources, which are taken from \citet{pestalozzi05}.} \\
(4){ The speculation about whether the CO gas is associated with the ${\rm CH_3OH}$ maser. The letter "q" stands for "questionable" (if the velocity difference between the maser source and the ${\rm ^{13}CO}$ line is in the range from 5 to 10 km s$^{-1}$, see Section 3.1).} \\
(5){ The kinematic distances of the maser sources. For the sources with an ambiguity, the two solutions (near and far distances) are both presented. If there are previous observational results helping to determine the distance, the reference number is given in square brackets. The recommended distance value is labeled with an asterisk.} \\
(6){ Other distance measurement. The reference number are given after the value.} \\
The indications for the reference numbers in column (5) and (6) are: [a1]\citet{urquhart11}, [a2]\citet{anderson09}, [b1]\citet{kuchar94}, [b2]\citet{roman09}, [b3]\citet{matthews09}, [b4]\citet{pandian08}, [c1]\citet{peretto09}, [c2]\citet{ragan12}, [c3]\citet{simon06}, [c4]\citet{he12}, [d1]\citet{faundez04}, [e1]\citet{zavagno02}, [e2]\citet{gwinn92}, [e3]\citet{bf82}, [e4]\citet{xu11}, [e5]\citet{sato10}, [e6]\citet{russeil03}, [e7]\citet{yang91}, [e8]\citet{hofner94}\\
The references are organized into five group according to the different features or procedures used to determine the distance (Section 3.3): \\
$[a1]$ and $[a2]$: 21-cm HI absorption towards background (UC) H{\sc ii} regions;\\
$[b1]$ to $[b4]$: 21-cm HI emission and self absorption;\\
$[c1]$ to $[c4]$: Infrared absorption features;\\
$[d1]$: An indication that the adopted distance should make the source closer to a spiral arm;\\
$[e1]$ to $[e8]$: Other distance measurements besides the kinematical method.\\
(7){ The distance value adopted for this paper. } \\
(8){ Galactocentric distance. } \\
(9){ The full-width half-maximum (FWHM) line width of the ${\rm ^{13}CO}~(1-0)$ line component from Gaussian fitting. } \\
(10){ The excitation temperature of the ${\rm ^{12}CO}~(1-0)$.} \\
(11){ The ${\rm H_2}$ column density, which is derived from the ${\rm ^{13}CO}~(1-0)$ line emission. } \\
\end{flushleft}
}
\end{landscape}
\end{longtab}
}

\centering
\begin{titlepage}
{\huge Online Material}
\end{titlepage}

\renewcommand\thetable{S\arabic{table}}
\setcounter{table}{0}

\onecolumn{
\begin{longtab}
\begin{landscape}
\small{                                                                                                                                                                             
\begin{longtable}{@{\extracolsep{\fill}}lllllll}                                                                                                                                             
\caption{${\rm ^{12}CO}$ and ${\rm ^{13}CO}$ Line Properties.}  \\                                                                      
\hline                                                                                                                                                                                     
Name               & R.A.               & Dec.              &  $T_{\rm mb}({\rm ^{12}CO})$     &  $T_{\rm mb}({\rm ^{13}CO})$     &  $\Delta V{\rm (^{13}CO})$   &  $\int T_{\rm mb}{\rm (^{13}CO)}{\rm dv}$   \\     
\quad              & (J2000)            & (J2000)           &  K                               &  K                               &  km s$^{-1}$                 &  (K km s$^{-1}$)                            \\     
(1)                & (2)                & (3)               & (4)                              &  (5)                             &  (6)                         &  (7)                                        \\     
\hline \endfirsthead
\caption{continued.}\\
\hline                                                                                                                                                                          
Name               & R.A.               & Dec.              &  $T_{\rm mb}({\rm ^{12}CO})$     &  $T_{\rm mb}({\rm ^{13}CO})$     &  $\Delta V{\rm (^{13}CO})$   &  $\int T_{\rm mb}{\rm (^{13}CO)}{\rm dv}$   \\     
\quad              & (J2000)            & (J2000)           &  K                               &  K                               &  km s$^{-1}$                 &  (K km s$^{-1}$)                            \\     
(1)                & (2)                & (3)               & (4)                              &  (5)                             &  (6)                         &  (7)                \\   
\hline \endhead                                                                                                                                                                   
8.68-0.36a          & 18:06:23.50       & -21:37:23.0       &  $20.7(1.45)$                   &  $18.8(1.32)$                    &   $03.88(0.15)$              &   $030.12(2.159)$    \\
8.68-0.36b          & 18:06:23.50       & -21:37:23.0       &  $20.7(1.45)$                   &  $12.7(0.89)$                    &   $04.46(0.27)$              &   $023.38(2.174)$    \\     
9.62+0.19           & 18:06:14.66       & -20:31:31.6       &  $13.2(0.92)$                   &  $09.2(0.64)$                    &   $05.44(0.10)$              &   $018.24(0.274)$    \\     
10.32-0.26          & 18:09:24.20       & -20:08:07.0       &  $17.8(1.25)$                   &  $07.0(0.49)$                    &   $06.24(0.19)$              &   $016.95(0.431)$    \\     
10.44-0.01          & 18:08:38.21       & -19:51:49.5       &  $13.4(0.94)$                   &  $06.2(0.43)$                    &   $07.58(0.33)$              &   $020.61(1.237)$    \\     
10.62-0.29          & 18:10:09.90       & -19:53:22.0       &  $25.8(1.81)$                   &  $11.3(0.79)$                    &   $04.55(0.04)$              &   $021.27(0.190)$    \\     
10.95+0.02          & 18:09:39.30       & -19:26:28.0       &  $12.9(0.90)$                   &  $04.5(0.32)$                    &   $03.85(0.57)$              &   $007.15(0.779)$    \\     
11.03+0.06          & 18:09:39.70       & -19:21:36.0       &  $07.3(0.51)$                   &  $03.8(0.27)$                    &   $04.05(0.19)$              &   $006.88(0.294)$    \\
\hline                                                                                                                                                                                     
\end{longtable}                                                                                                                                                                            
}
\end{landscape}
\end{longtab}                                                                                                                                                                                          
\clearpage

\setcounter{table}{1}

\begin{table*}
{\small
\caption{Physical parameters and radio continuum emission of the identified cores.}
\begin{tabular}{llllll}
\hline
Source             &  $T_{\rm ex}$        &  $N_{\rm H_2}$            &  $R_{\rm core}$   &  $F_{\rm cm}$          &  IR emission  \\
\quad              &  (K)                 &  ($10^{22}$ cm$^{-2}$)    &  (pc)             &  (mJy)                 &  \quad        \\
(1)                &  (2)                 &  (3)                      &  (4)              &  (5)                   &  (6)          \\
\hline                                                                             
G008.68-0.37       &  $24.1(1.7)$         &  $07.77(02.33)$           &  2.2              &  $<2.3^m$              &  p,d     \\
G010.47+0.03       &  $16.8(1.2)$         &  $02.15(00.65)$           &  1.8              &  $65.52^m$             &  d       \\
G010.62-0.38       &  $29.3(2.1)$         &  $03.05(00.92)$           &  1.2              &  $1899.45^m$           &  d       \\
G012.03-0.04       &  $17.5(1.2)$         &  $01.91(00.57)$           &  1.6              &  $<2.3^m$              &  p       \\
G012.79-0.19       &  $25.2(1.7)$         &  $29.16(08.73)$           &  1.6              &  $727.65^m$            &  p,d     \\
\hline
\end{tabular} \\
}
{\small
\begin{flushleft}
The description for the columns:\\
(1){ The source names, which are also their Galactic coordinates. If one region has multiple ${\rm ^{13}CO}$ cores, they are given suffix number "(1), (2), (3)...". } \\
(2){ The excitation tempearture of the ${\rm ^{12}CO}~(1-0)$ at the emission peak of the core. } \\
(3){ The ${\rm H_2}$ column density at the emission peak of the core. } \\
(4){ The effective radius of the core, which is calculated from using $\pi R_{\rm core}^2=A_{\rm core}$ where $A_{\rm core}$ is the area encircled by the 50 \% contour in the ${\rm ^{13}CO}~(1-0)$ map (in unit of pc$^2$).} \\
(5){ The flux of the radio continuum emission associated with the core. The values are extracted from several works, including the 6 cm $^m$MAGPIS survey \citep{helfand06}, and the observations of $^g$\citet{gregory91}, $^r$\citet{rodriguez81}, $^k$\citet{kurtz94}, and $^w$\citet{walsh98}. The observational wavelength is $\lambda=6.1$ cm for \citet{gregory91} and \citet{rodriguez81}, $\lambda=3.5$ cm for \citet{kurtz94}, and $\lambda=3.5$ cm for \citet{walsh98}. The detection limit is $\sigma\simeq2.3$ mJy for MAGPIS and $\sigma\simeq6$ mJy for \citet{gregory91}.} \\
(6){ Infrared emission feature associated with the $^{13}$CO cores as shown in Figure S1. For the cores associated with the IR emission, the emission features are classified as "point-like" (p) and "diffused" (d). }\\
\end{flushleft}
}
\end{table*}
\clearpage 

\begin{table*}
\caption{Infall velocities and mass infall rates. \label{tbl_s3}}
\begin{tabular}{lccc}
\hline\hline
\quad                 &  $V_{\rm in}^a$   &  $n({\rm H_2})^b$     &  $\dot{M}_{\rm inf}^c$               \\
Source Name           &  (km s$^{-1}$)    &  ($10^4$ cm$^{-3}$)    &  ($10^{-3}~M_{\odot}$ year$^{-1}$)  \\
\hline
G08.68-0.37           &  $0.7\pm0.2$      &  1.5                   &  $16\pm5$          \\
G12.79-0.19           &  $1.7\pm0.4$      &  3.0                   &  $20\pm4$          \\
G35.20-1.73(1)        &  $0.4\pm0.1$      &  1.0                   &  $2\pm0.5$         \\
G80.85+0.43(1)        &  $0.3\pm0.1$      &  0.9                   &  $1.6\pm0.5$       \\
\hline
\end{tabular} \\
{\small
$a.${ Infall velocities derived from the double-peaked $^{12}$CO profiles.}  \\
$b.${ Ambient densities calculated from $N({\rm H_2})/2R_{\rm core}$.} \\
$c.${ Mass infall rate.} \\
}
\end{table*}
\clearpage

\renewcommand\thefigure{S\arabic{figure}}
\setcounter{figure}{0}
\begin{figure*}
\centering
\includegraphics[angle=0,width=0.9\textwidth]{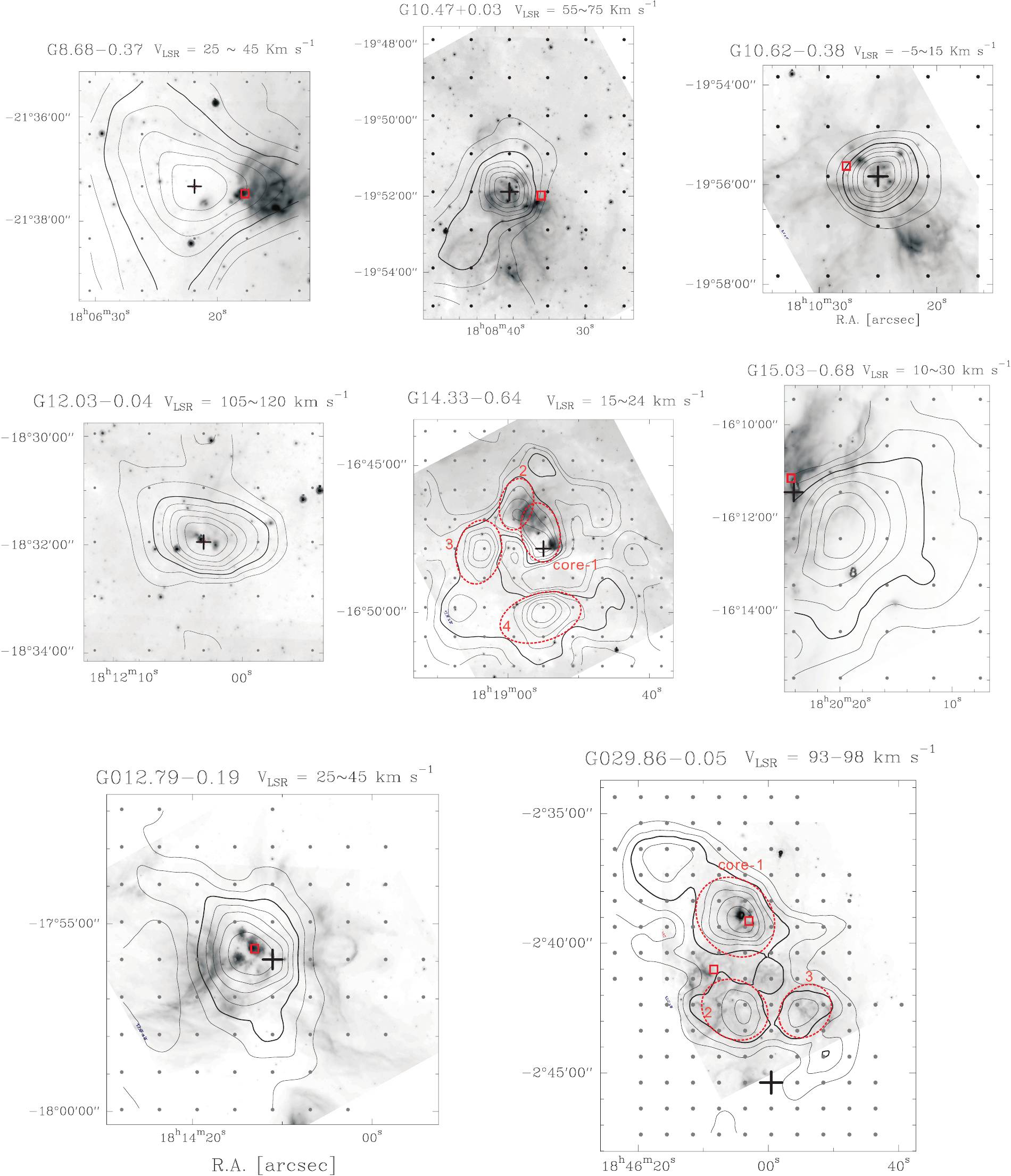} \\
\caption{\small The $^{13}$CO $(1-0)$ maps (contours) overlaid on the far-infrared images (gray). The thick contour in each map denote the 50 per cent level of the maximum intensity. For each map, the interpolation is performed upon a grid with $100\times100$ pixels. For all the regions, it is guaranteed that the pixel size is far smaller than the mapping step. And in each panel, the black cross indicates the central position (where the maser locates), the square indicates the radio continuum source (plausible Ultra- or Hyper-Compact H{\sc ii} regions, if exists), the dashed ellipses indicate the identified cores if there are overlapped multiple cores. }
\end{figure*}
\clearpage

\setcounter{figure}{0}
\begin{figure}
\centering
\includegraphics[angle=0,width=0.9\textwidth]{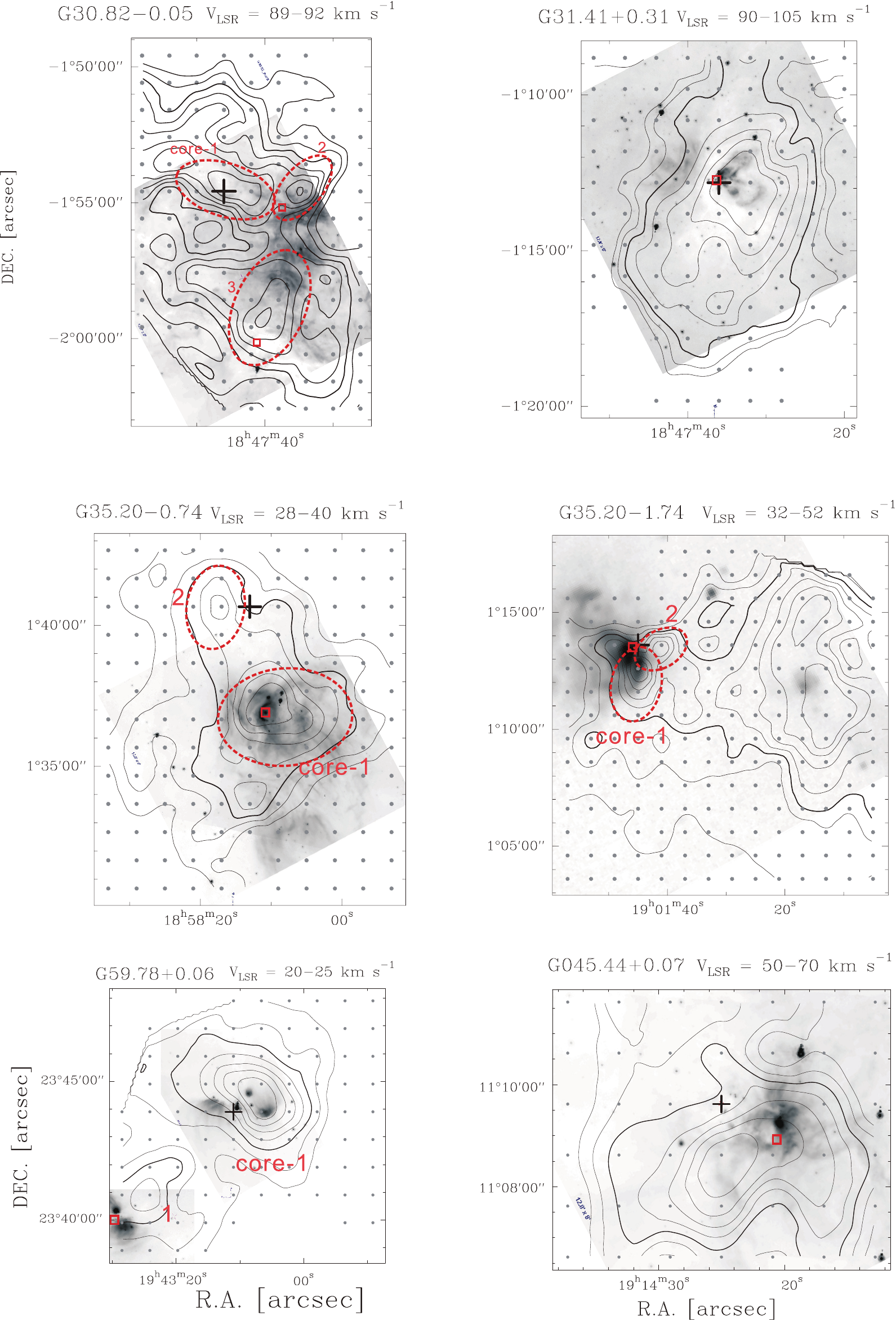} \\
\caption{\small continued.}
\end{figure}
\clearpage

\setcounter{figure}{0}
\begin{figure}
\centering
\includegraphics[angle=0,width=0.9\textwidth]{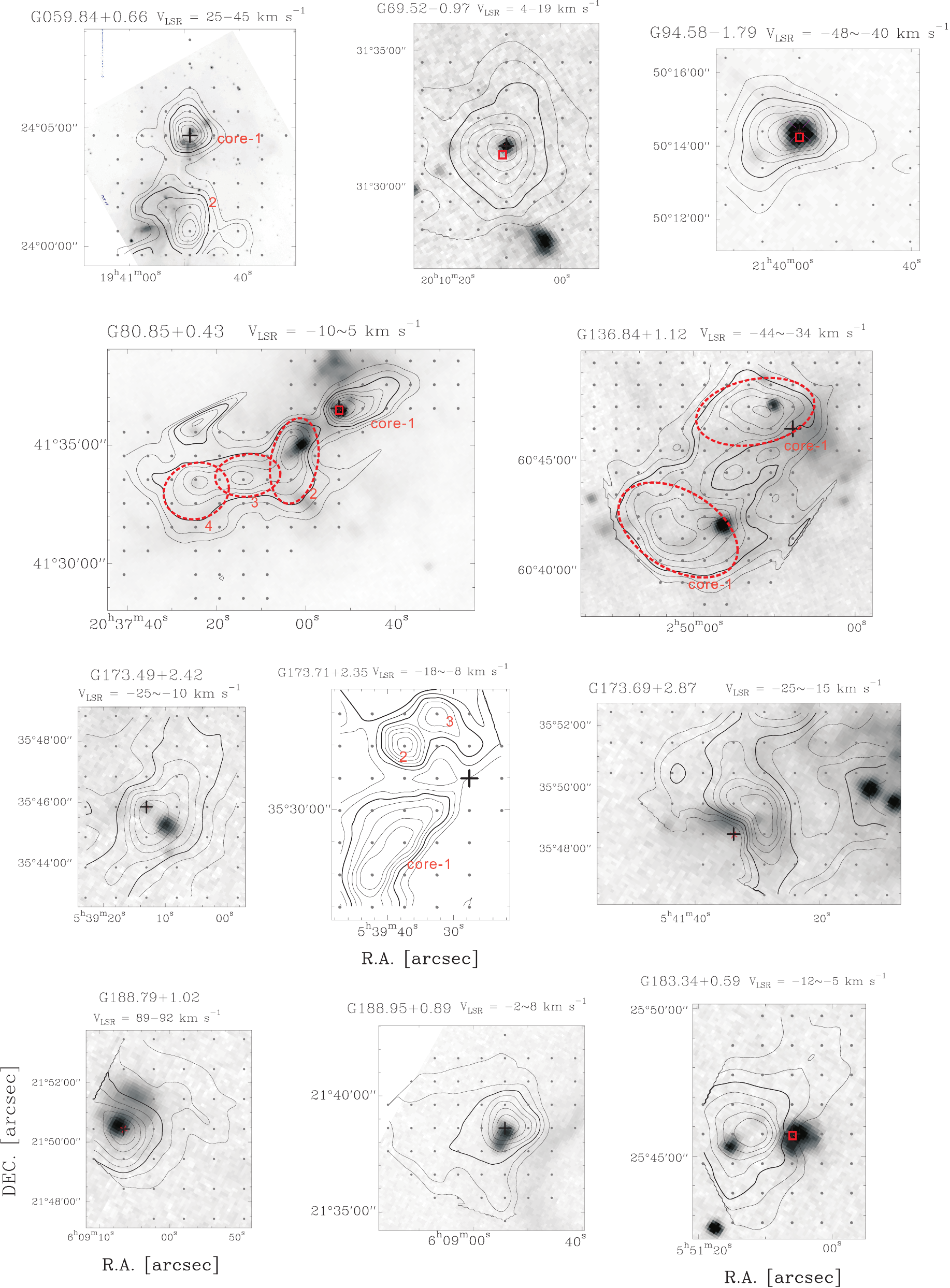} \\
\caption{\small continued.}
\end{figure}
\clearpage

\setcounter{figure}{0}
\begin{figure}
\centering
\includegraphics[angle=0,width=0.9\textwidth]{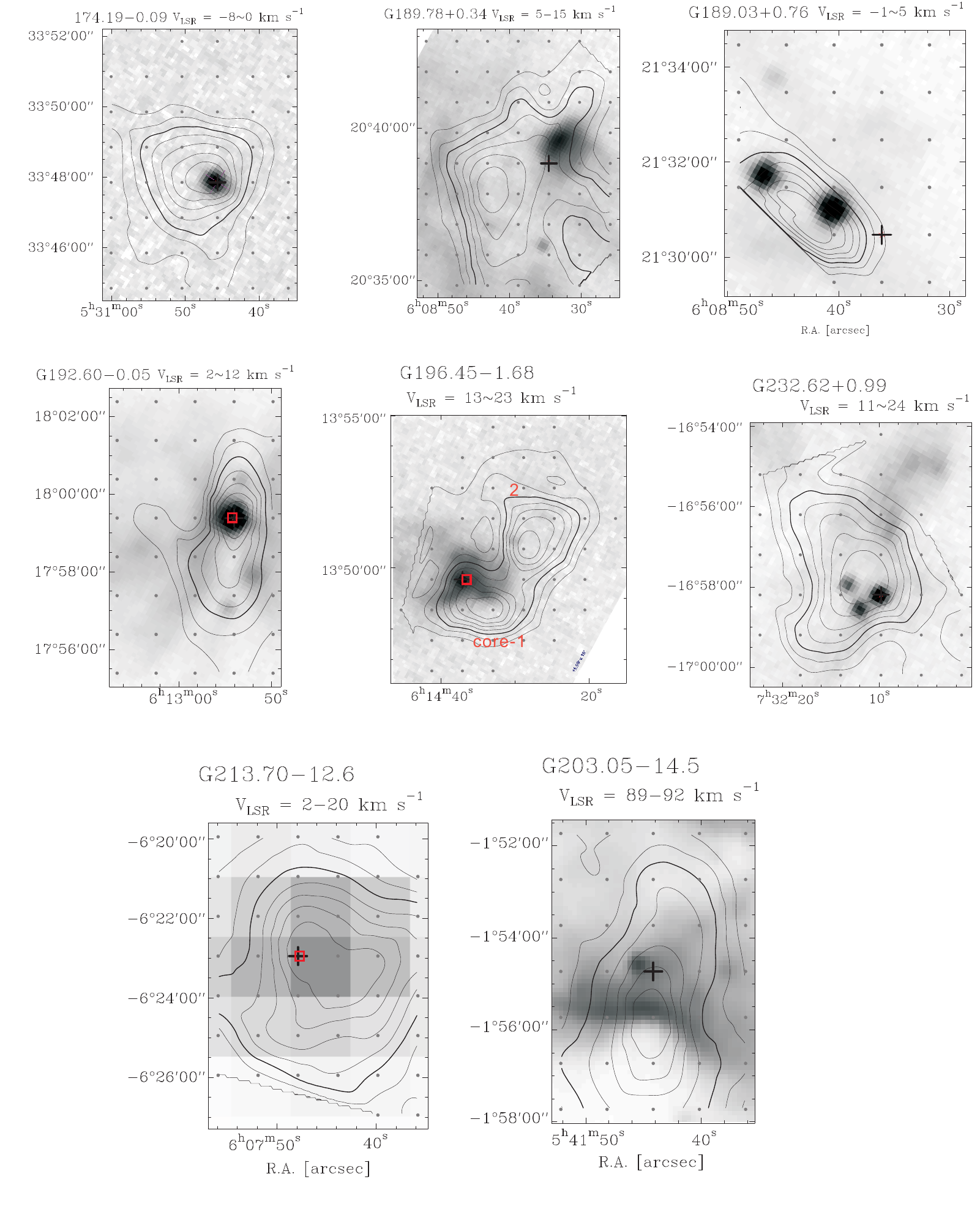} \\
\caption{\small continued.}
\end{figure}

\begin{figure}
\centering
\includegraphics[angle=0,width=0.8\textwidth]{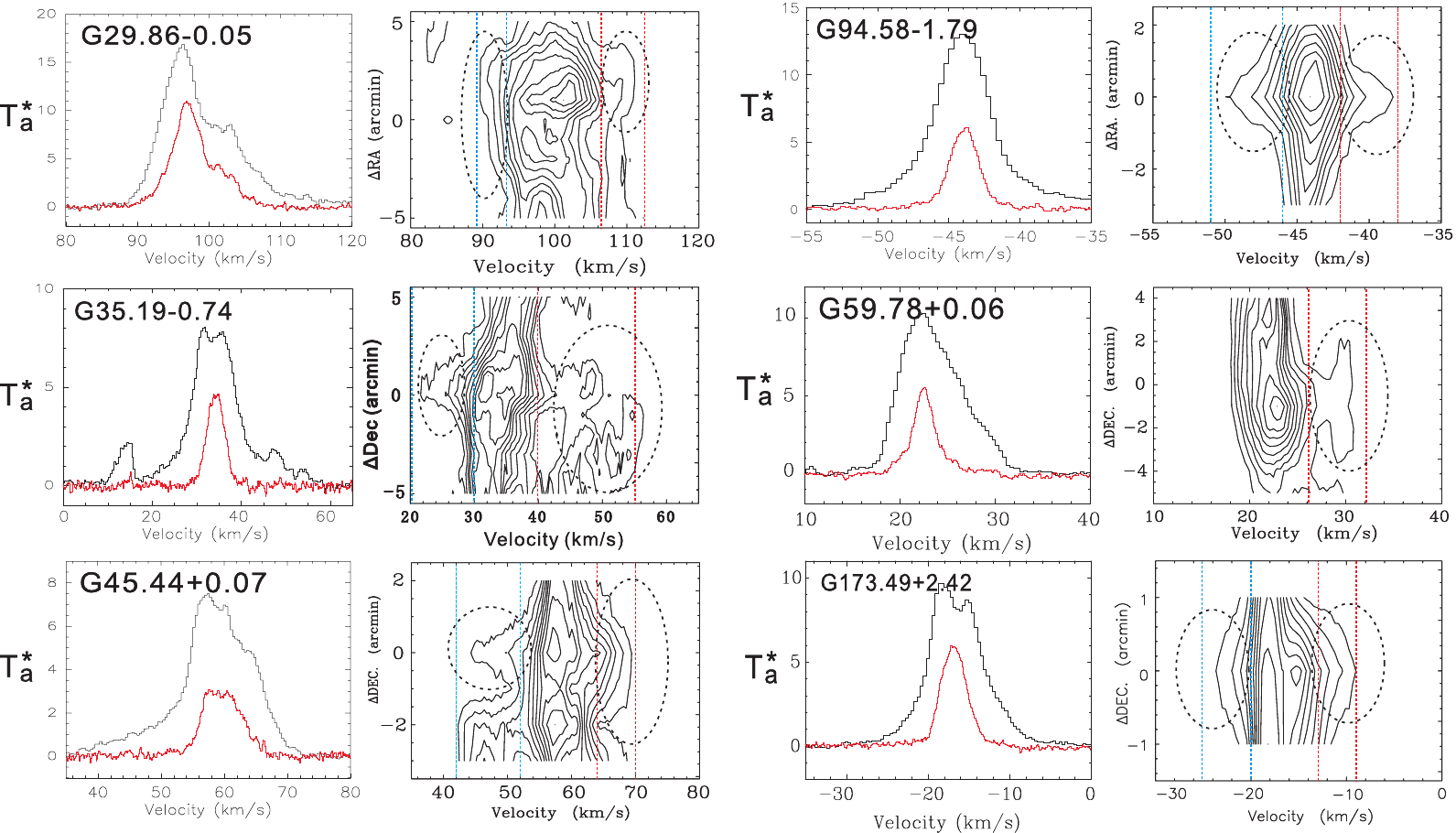} \\
\caption{(a) The CO spectra (${\rm ^{12}CO}$ in black and ${\rm ^{13}CO}$ in red) at the mapping point closest to the core center and position-velocity diagrams of the outflow candidates. The dashed lines indicate the velocity ranges adopted as the line wings for the four sources (G59.78 only has evident red wings). The swelled line wings that may indicate the outflow are also encircled with the dashed ellipses.}
\end{figure}

\setcounter{figure}{1}
\begin{figure}
\centering
\includegraphics[angle=0,width=0.7\textwidth]{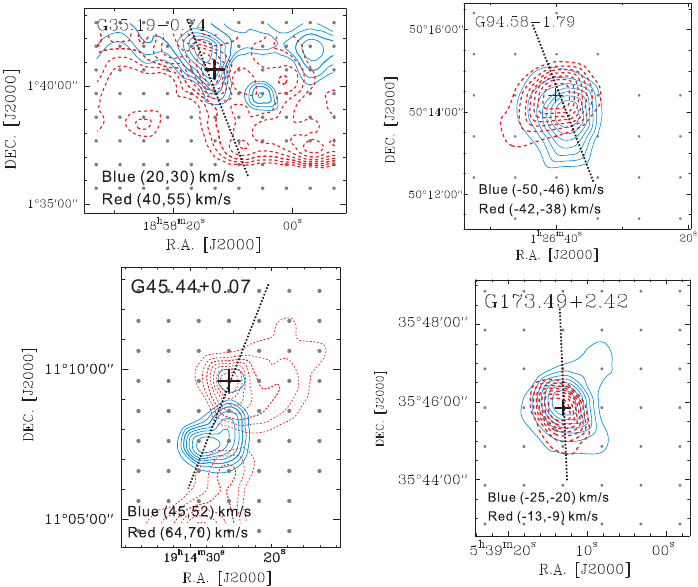} \\
\caption{(b) The outflow emissions. The blue solid and red dashed contours indicate the blue- and red-shifted emissions, respectively. The dashed line in each panel indicates the speculated outflow orientation. The velocity ranges for the integration are shown in each panel. The black cross indicates the central position (maser source).}
\end{figure}

\begin{figure}
\centering
\includegraphics[angle=0,width=0.5\textwidth]{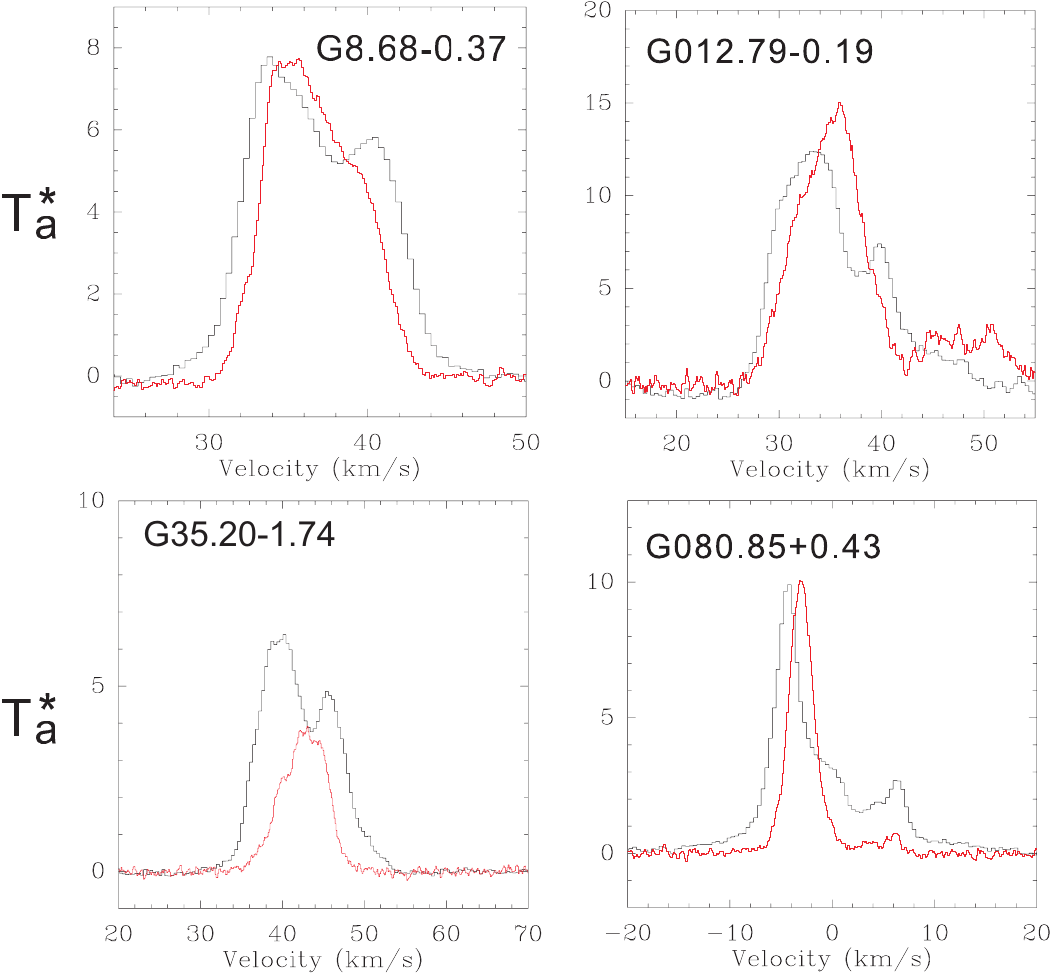} \\
\caption{The CO spectra of the infall candidates. The spectra of another infall candidate. }
\end{figure} 
}

\begin{figure}
\centering
\includegraphics[angle=0,width=0.8\textwidth]{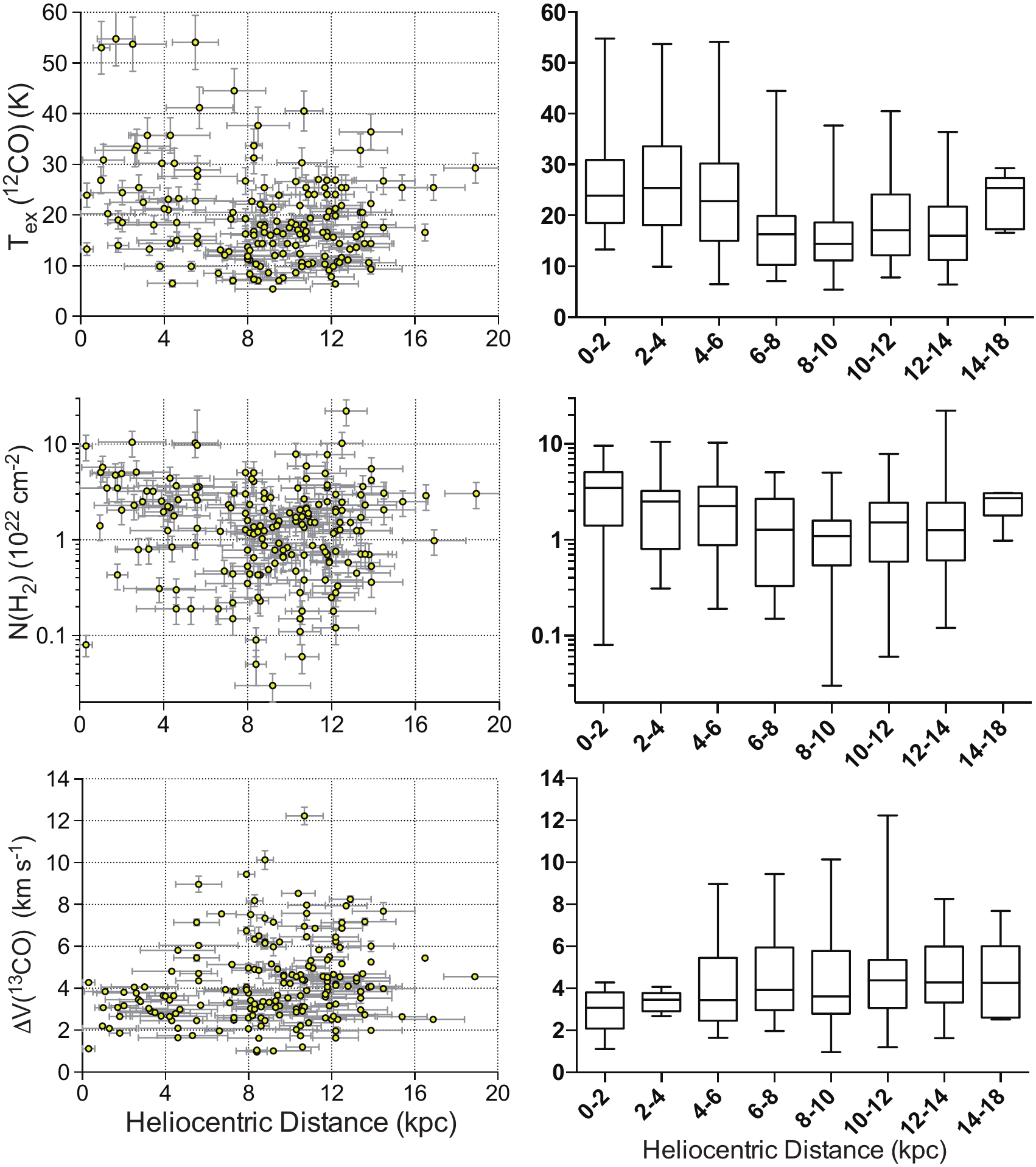} \\
\caption{Same as Figure 1 but with the far distance adopted for all the sources with two $D_{\rm kin}$ solutions. The statistical box is the same as described in Figure 3.}
\end{figure}

\end{document}